# Nanoscale Mapping of Transition Metal Ordering in Individual $LiNi_{0.5}Mn_{1.5}O_4$ Particles Using 4D-STEM ACOM Technique

Gozde Oney*[1,2,3 #], Fayçal Adrar[2,3 #], Junhao Cao[2,3], Chunyang Zhang[4], Muriel Véron[4], Matthieu Bugnet[7], Emmanuelle Suard[5], Jacob Olchowka[1,3,6], Laurence Croguennec[1,3,6], François Weill*[1,3], Arnaud Demortière*[2,3,6]

[#] equal contributions

[1] CNRS, Univ. Bordeaux, Bordeaux INP, ICMCB UMR 5026, F-33600 Pessac, France
[2] Laboratoire de Réactivité et de Chimie des Solides (LRCS), CNRS-UPJV UMR 7314, Amiens, France
[3] Réseau Français sur le Stockage Electrochimique de l'Energie (RS2E), FR CNRS 3459, Amiens, France
[4] Université Grenoble Alpes, CNRS, Grenoble INP, SIMaP, F-38000 Grenoble, France
[5] Institut Laue-Langevin (ILL), BP 156, 71 Avenue des Martyrs, 38042 Grenoble, France
[6] ALISTORE-ERI European Research Institute, FR CNRS 3104, Amiens, France
[7] CNRS, INSA Lyon, Université Claude Bernard Lyon 1, MATEIS, UMR 5510, 69621 Villeurbanne, France

**Corresponding Authors:**
francois.weill@icmcb.cnrs.fr, arnaud.demortiere@cnrs.fr, gozde.oney@cea.fr

**Keywords:** Spinel $LiNi_{0.5}Mn_{1.5}O_4$, ordered/disordered phases, 4D-STEM technique, Electron microscopy, phase identification at particle level.

**Abstract:** The electrochemical performance of the spinel $LiNi_{0.5}Mn_{1.5}O_4$, a high-voltage positive electrode material for Li-ion batteries, is influenced by the transition metal arrangement in the octahedral network, leading to disordered ($Fd\bar{3}m$ S.G.) and ordered ($P4_332$ S.G.) structures. However, widely used techniques lack the spatial resolution necessary to elucidate the ordering phenomenon at the particle scale. Using the 4D-STEM technique, we present the first direct observation of ordering distribution in individual $LiNi_{0.5}Mn_{1.5}O_4$ particles with nanometric spatial resolution. We propose a quantification method for the local degree of ordering based on the ratio of ordered to disordered spinel lattices along the particle thickness extracted from electron diffraction spot intensities. In an ordered spinel $LiNi_{0.5}Mn_{1.5}O_4$, the transition metal ordering is consistently observed throughout the primary particle. However, the extent of ordering in the spinel phase depends on its distribution at the particle scale, a factor influenced by the annealing conditions. The 4D-STEM analysis elucidates the boundary between highly-ordered and low-ordered $LiNi_{0.5}Mn_{1.5}O_4$ particles.

## Introduction

The increasing demand for electric vehicles and portable devices has driven the need to develop low-cost and higher energy density Li-ion batteries (LIBs).[1] Among the eligible candidates for positive electrode materials, the spinel $LiNi_{0.5}Mn_{1.5}O_4$ (LNMO) with a Co-free, Li-poor, and Mn-rich composition allows to reduce the overall cost[2]. It also delivers a high energy density of 650 Wh/kg thanks to its high working potential of 4.7 V vs. $Li^+$/Li and a theoretical capacity of 147 mAh/g.[3,4]

Research on LNMO has shown that the degree of ordering between nickel and manganese atoms in the spinel framework has a major effect on its electrochemical performance. This degree of ordering is mainly driven by the Ni/Mn stoichiometry and annealing conditions.[5–8] The disordered LNMO materials are often characterized by a Ni/Mn ratio lower than 1/3, and thus by the presence of a mixed valence for Mn ($Mn^{3+}$, $Mn^{4+}$) for charge compensation. In this case, the Ni and Mn transition metals are statistically distributed on the 16d octahedral sites within a unit cell described in the $Fd\bar{3}m$ space group (Fig. 1a). Disordered spinels typically demonstrate electrochemical capacities near the theoretical limit. However, a portion of the attainable capacity is impeded by the presence of electrochemically inactive impurities. Concurrently, these impurities are susceptible to degradation and capacity decay through Mn dissolution, attributable to the mixed valence states of $Mn^{3+}$ and $Mn^{4+}$.[9] On the contrary, ordered LNMO is described within the $P4_332$ space group with Ni and Mn atoms in distinct octahedral crystallographic sites, 4b and 12d, respectively (Fig.1e). Its lower specific capacity, especially at high rates, is attributed to a weaker electrical conductivity due to an extended order between $Ni^{2+}$ and $Mn^{4+}$ cations. Its superior structural stability during extended cycling is attributed to the absence of $Mn^{3+}$ redox activity in this perfectly stoichiometric composition with a Ni:Mn ratio of 1:3.[10]

Analytical techniques, such as neutron diffraction, Raman and NMR spectroscopies, reveal how extended is the transition metal ordering in spinel LNMO[4,7,11–15]. To explain the underlying ordering mechanism, an antiphase boundary model has been proposed based on averaged structural information obtained from neutron diffraction data collected for different LNMO compositions.[12,16] According to this model, the LNMO sample consists solely of coherent ordered domains that are increasing in size with the degree of ordering rather than of mixtures of ordered and disordered domains. To validate or refute this model, complementary localized observations and detailed investigations are essential.

In 2022, a spatially-resolved d-spacing and strain mapping analysis was performed by Spence et al.[17] on a partially ordered LNMO sample using hard X-ray nano-diffraction with 50 nm of spatial resolution. It revealed a heterogeneous distribution of *d*-spacing within the selected particles, which was interpreted as domains with different $Mn^{3+}$ contents (*i.e.* with different Ni/Mn ratios), and thus heterogeneous degrees of ordering. Nevertheless, the direct correlation between the lattice parameter and the composition remains inadequately defined for this system. Additionally, the presence of impurities can strongly influence the actual composition and thus lattice parameters of the spinel phases.[14,18,19] Additionally, it is noteworthy that the literature reports surprisingly $Mn^{3+}$-rich ordered and $Mn^{3+}$-poor disordered LNMO samples[8,13] , underlying the complexity of the ordering-composition relationship.

More recently, Tertov et al.[35] combined neutron diffraction with high-resolution scanning transmission electron microscopy (HR-STEM) to show that ordering in LNMO is not uniform, but instead forms a mosaic of nanosized ordered domains separated by disordered regions. They further established a correlation between Mn content and the degree of ordering, with higher Mn fractions favoring disordering. These results provide valuable nanoscale insights into the link between composition and ordering, obtained with a depth resolution of 10 nm. At the same time, questions remain about how ordering heterogeneity extends throughout the full particle thickness and how it can be quantified with high spatial resolution, motivating further investigation.

Transmission electron microscopy (TEM) achieves exceptionally high spatial resolution due to the emergence of spherical aberration correctors. Recently, the four-dimensional scanning transmission electron microscopy (4D-STEM) has emerged as a powerful technique, enabling the acquisition of spatially resolved datasets that combine two-dimensional (2D) electron diffraction patterns in reciprocal space with 2D real-space scanning coordinates. The use of a precessed electron beam further improves diffraction pattern quality, facilitating the generation of high-resolution structural maps. In materials science, 4D-STEM has been effectively employed to produce orientation and phase maps[20–22], strain maps[23,24], and ptychography reconstructed imaging [25,26]. One widely adopted implementation is the Automated Crystal Orientation Mapping technique (ACOM)[27–29], developed by Rauch *et al.* and commercialized by Nanomegas (Astar), which offers spatial resolution down to 1 nm and has been extensively used for metals and alloys[30]. In parallel, open-source Python-based toolkits such as *py4DSTEM*[31] and *pyXEM*[32] have been developed to support advanced data analysis in 4D-STEM. More recently, the *ePattern* tool introduced by Folastre et al.[33] has further improved pattern matching by enabling precise peak detection and robust denoising, thereby increasing the accuracy and reproducibility of orientation and phase determinations. Notably, *ePattern* also reduces user bias during data processing, ensuring more consistent and reliable interpretation of complex 4D-STEM datasets[33b.]

In this study, we introduce the application of the 4D-STEM technique to determine the extension of the transition metal ordering in spinel LNMO and to compare this result with analyses performed by neutron powder diffraction. This study enables a statistically robust, spatially resolved characterization of polycrystalline LNMO powders by combining 4D-STEM techniques with a data science approach. Individual particles are probed across the sample, allowing for the extraction of statistically meaningful information beyond averaged measurements. We also comprehensively discuss the experimental parameters and data analysis workflows used to interpret the 4D-STEM datasets. Crucially, we demonstrate for the first time that the degree of transition metal ordering in spinel LNMO can be quantitatively mapped at the single-particle level. Our approach discriminates between perfectly ordered, partially ordered, and disordered regions within individual crystallites.

## Description of the global structure

This study focuses on three LNMO samples obtained through different annealing conditions from the same initial batch that targeted a stoichiometric Ni:Mn 1:3 ratio, each showing particles with platelet-like morphology (180 nm in average thickness and 1 µm in average diameter) and having different degrees of transition metals' ordering as determined by average (*i.e.* at long range distances) analysis. Preparation of the samples are discussed in detail in supplementary information. As shown by the X-ray and neutron diffraction patterns given in Fig. S1, all samples appear as single spinel phases with cubic structures and lattice parameters ranging between 8.1759(2) Å for the disordered LNMO (*d*-LNMO), 8.1697(6) Å and 8.1679(4) Å for the two ordered LNMO's (*o1*-LNMO and *o2*-LNMO, respectively). The decrease of the cell parameter within this series is attributed to changes in Ni/Mn composition and thus in $Mn^{3+}$ content: the richer in $Mn^{3+}$ cations, the higher is the cell parameter of the spinel phase. The neutron powder diffraction allows to characterize the extent of ordering in LNMO. The neutron diffraction pattern of disordered *d*-LNMO can be indexed in the $Fd\bar{3}m$ space group, while the other two collected patterns for *o1*-LNMO and *o2*-LNMO were indexed in the $P4_332$ space group. The ordering peaks' width was used to determine the size of the ordered domains within the spinel structure. For *o1*-LNMO, the ordered domain size was calculated to be 33 nm, while *o2*-LNMO presented larger domains of 45 nm. The composition of the spinel phase was calculated by refining the neutron powder diffraction patterns with Rietveld method and found to be $LiNi_{0.462}Mn_{1.538}O_4$ for *d*-LNMO, $LiNi_{0.476}Mn_{1.524}O_4$ for *o1*-LNMO, and $LiNi_{0.492}Mn_{1.508}O_4$ for *o2*-LNMO as reported in Fig. S2 and Table S1. The estimated $Mn^{3+}$ content required for charge compensation is 4.9 % in the disordered *d*-LNMO sample, and decreases to 1.1 % in the highly ordered *o2*-LNMO spinel. Interestingly, the *o1*-LNMO sample, despite its classification as ordered, exhibits a slightly Mn-rich composition, with an estimated $Mn^{3+}$ content of 3.2 %, which is atypical for a stoichiometrically ordered phase.

Upon examining the electrochemical signatures obtained for the three LNMO materials, as depicted in Fig. S3, distinct behaviors are observed between these three samples. In agreement with their structural descriptions using Neutron Powder Diffraction (NPD), the $Ni^{2+}/Ni^{3+}$ redox couple activity is associated to a solid-solution mechanism for *d*-LNMO, as expected for disordered LNMO, whereas a biphasic reaction plateau is observed for the two *o1*- and *o2*-LNMO samples, as expected for ordered LNMO. However, a notable discrepancy arises in the contribution of the Mn redox reaction occurring at lower potentials. The $Mn^{3+}$ content is estimated to be 4 % in disordered *d*-LNMO. As expected, this content decreases to 1 % in the sample ordered *o2*-LNMO. However, even though in agreement with the NPD refinement results, *o1*-LNMO showcases 4 % Mn redox activity and also presents two-shouldered redox signatures at 3.96 and 4.20 V, as opposed to a single step around 4.10 V for *d*-LNMO and *o2*-LNMO. This behavior is seen for Mn-rich ordered spinels before and attributed to a Ni/Mn disordering between 4b and 12d sites.[8,35]This highlights that even though both *o1*- and *o2*-LNMO are ordered according to their description in the $P4_332$ space group thanks to NPD, they significantly differ in terms of their compositions and local arrangements of their transition metals leading to the presence of defects.

## (Dis)Ordering with spatial resolution

For each prepared LNMO sample, we have examined by 4D-STEM ACOM technique individual particles, selected randomly according to their position on the TEM grid. A single isolated particle is chosen (Fig. 1(b-f-i)) for each sample to illustrate the mapping of dis(ordering) at the particle level.

Using the 4D-STEM technique, we performed full-particle scans in diffraction mode with a quasi-parallel electron beam, employing a convergence semi-angle of 0.4 mrad and a step size of 10 nm. A precession angle of 0.8° was applied to mitigate dynamical scattering effects and enhance pattern quality. This configuration enabled the acquisition of high-resolution diffraction datasets within a reasonable experimental duration (~30 minutes per map), while also streamlining subsequent data processing. At each pixel (10 × 10 nm²), the recorded diffraction patterns were systematically compared using pattern matching to simulated reference patterns corresponding to the two distinct structural phases of $LiNi_{0.5}Mn_{1.5}O_4$: the fully disordered spinel phase ($Fd\bar{3}m$ space group) and the perfectly ordered spinel phase ($P4_332$ space group). These phases can be distinguished based on their electron diffraction signatures, notably the presence or absence of reflections with mixed-parity (hkl) indices, which are permitted in the primitive $P4_332$ lattice but forbidden in the face-centered $Fd\bar{3}m$ lattice, as illustrated in Fig. S4.

Fig. 1(c-g-j) shows the electron diffraction pattern taken at the selected pixel indicated with a cross in Fig. 1(b-f-i). Additionally, Fig. 1(d-h-k) shows the line integration profile associated with the dashed-lined orientation given in the diffraction pattern. In all of them, the [110] direction is chosen in order to be able to distinguish the diffraction spots associated with the ordering (*i.e.* 110, 330, 550). The observed diffraction patterns and associated integration line profiles align perfectly with the phase mapping results (Fig. S5). In the pattern collected for the disordered *d*-LNMO the ordering spots are absent, whereas in the two ordered *o1*-LNMO and *o2*-LNMO patterns the ordering spots are clearly visible.

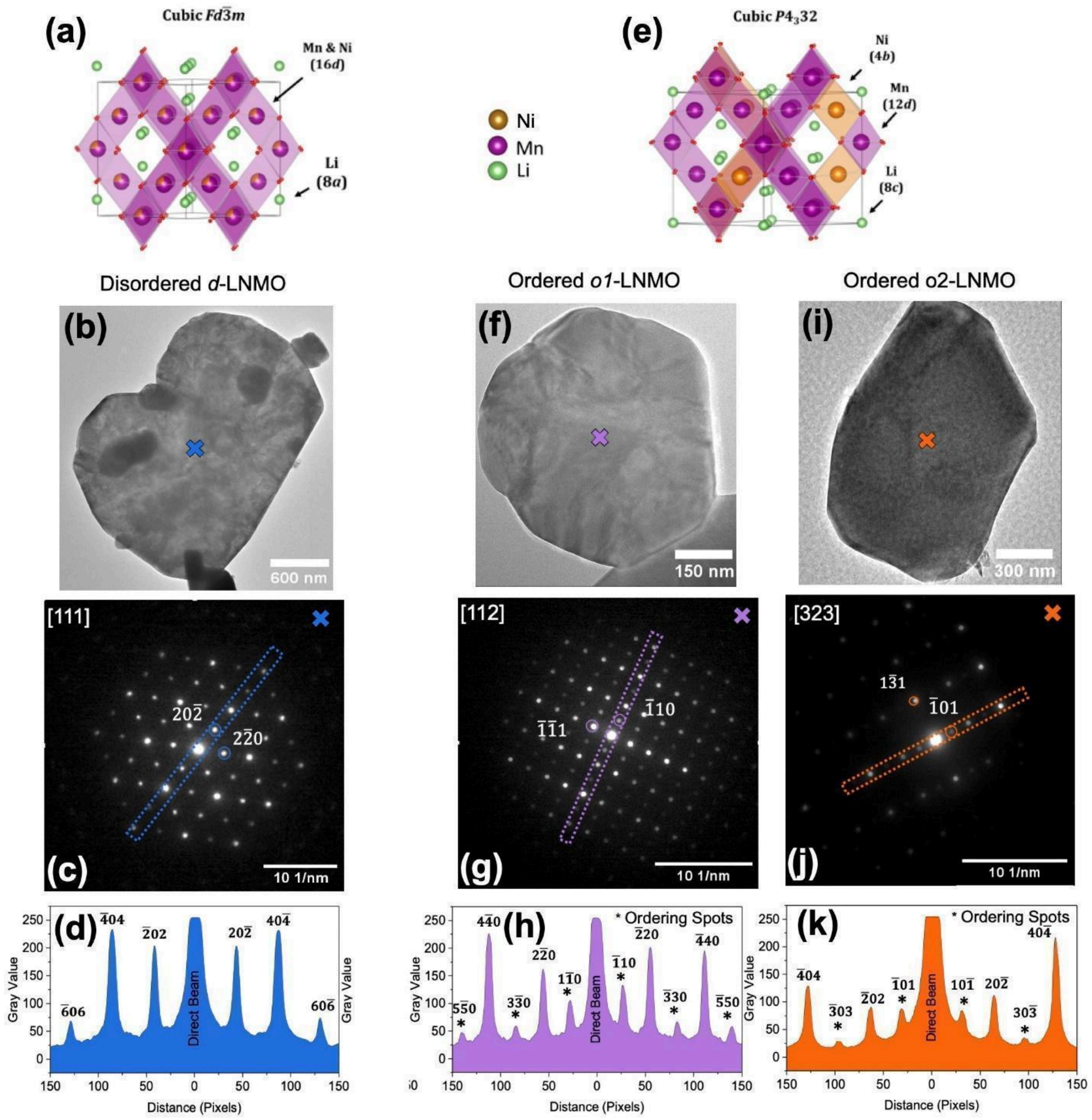

**Figure 1. Electron diffraction identifying ordered and disordered LNMO phases. (a,e)** Crystal structures of disordered LNMO described in the $Fd\bar{3}m$ space group and ordered LNMO described in the $P4_332$ space group, respectively. Lithium atoms are represented by green balls, manganese by purple balls, nickel by orange balls, and oxygen by red balls. The, below, from left to right, the *disordered (d-),* the *ordered (o1-)*, and the *ordered (o2-)* LNMO samples: **(b,f,i)** TEM images of selected particles for phase mapping. The colored crosses correspond to the electron diffraction patterns shown in (**c,g,j**). **(c-g-j)** Electron diffraction patterns acquired through 4D-STEM ACOM on the points localized by colored stars at the center of the particles in **(b,f,i)**. **(d,h,k)** Integration of line profiles on the dash lined orientation given in the associated diffraction patterns. Asterisks mark the ordering spots.

## Ideal phase mapping conditions

When performing phase mapping, careful selection of the particles to be analyzed is crucial to ensure the reliability of the results and to minimize the risk of misinterpretation or overestimation. Our detailed evaluation of the collected phase maps has revealed several critical parameters that must be taken into account to ensure accurate analysis. These parameters can be classified into two main categories: experimental constraints, such as specimen thickness and crystal orientation, and software-related limitations, including pattern matching accuracy and detection thresholds. Addressing both sets of factors is essential to obtain robust and interpretable spatial phase distribution maps.

The first essential experimental condition to be considered is the particle thickness. This is because the experimental diffraction patterns are compared with the theoretical ones calculated under kinematical considerations, where the incident wave maintains a constant amplitude upon scattering. This consideration can be experimentally valid when the specimen is ultra-thin (less than 20 nm). However, if the specimen is electron-transparent but not sufficiently thin, the incident electrons interact strongly with matter, and the diffracted beam can act as new multiple incident beams, leading to multiple scattering. To minimize these dynamical effects, we targeted a thin platelet morphology (180 nm in average thickness) and applied a beam precession of 0.8° throughout the mappings. However, when the particle thickness remains substantial, several artifacts may arise. In ordered phases, the intensity of characteristic ordering reflections may be significantly attenuated or even suppressed, leading to their non-detection. Conversely, in disordered regions, diffuse scattering can produce spurious intensity at positions corresponding to ordering reflections, potentially resulting in misidentification and inaccurate phase mapping (Fig. S6).

The second critical experimental parameter is the crystal orientation during data acquisition. Accurate identification of ordering reflections is significantly facilitated when the crystal is aligned along a zone axis, as this geometry optimizes the visibility and symmetry of characteristic diffraction spots. However, even a slight misorientation can lead to mistaken phase attribution. This sensitivity is illustrated in Fig. S7, and further evidenced by the deviation from the zone axis in the *o1*-LNMO particle, as revealed by Non-Negative Matrix Factorization (NMF) clustering[34] (Fig. S8). The underlying cause of this limitation stems from the pattern-matching algorithm employed in the ASTAR software, which determines the crystallographic phase by summing the positions and intensities of diffraction spots within a defined search radius (see Fig. S9). Deviations from ideal orientation can distort spot intensities and spatial distributions, thereby compromising the reliability of phase identification.

Even though all the ideal experimental conditions of phase mapping are achieved, it is possible to lose information due to software-related limits. One of the key factors influencing phase detection is the size of this radius used for matching experimental and theoretical patterns, as just mentioned. Due to the curvature of Ewald's sphere, when we move away from its center, the sphere intersects only a minor portion of the diffracting domains. Consequently, diffraction spots farther away from the central beam (corresponding to large g vector distances in reciprocal space) will have less intensity, especially for extra spots exhibiting much lower intensity. Since the software determines the phase based on a comparison of intensities at specific spot positions, selecting a too large search zone can lead to a shift in the choice of the software. For example, selecting the disordered phase instead of the ordered phase despite the presence of ordering spots (Fig. S9).

A second significant software-related limitation lies in the intensity detection threshold, which may result in phase misclassification or the designation of certain regions as unreliable (Fig. S10). This limitation stems from the reliance on theoretical intensity calculations based on single-crystal models of the $P4_332$ and $Fd\bar{3}m$ space groups under kinematical approximation. In contrast, experimental datasets often reflect the superposition of multiple stacked lattices, where dynamical scattering effects, though mitigated by beam precession, remain non-negligible and lead to unpredictable variations in diffraction spot intensities. As a result, regions with inherently weak or attenuated signals may be incorrectly assigned by the software, highlighting the need for cautious interpretation of low-intensity features in 4D-STEM datasets.

## Mapping the degree of ordering

Under the kinematical approximation, a correlation can be established between the intensity of ordering reflections in patterns and the number of ordered lattice planes contributing to the signal. In an ideal scenario, where dynamical scattering effects are negligible or effectively suppressed by beam precession, and the LNMO platelets exhibit uniform thickness, a fully ordered crystal with $P4_332$ symmetry is expected to produce an intensity ratio between ordering-specific reflections and those common to both ordered and disordered ($Fd\bar{3}m$) phases that closely aligns with theoretical predictions. However, if the experimentally measured intensity ratio of the ordering reflections is lower than expected, it suggests a partial or incomplete ordering along the beam direction. This deviation reflects a coexistence of ordered ($P4_332$) and disordered ($Fd\bar{3}m$) domains along the crystal thickness, leading to a cumulative diffraction intensity that underrepresents the ideal ordering contribution.

Since the phase mapping outputs showed overall $P4_332$ phase indexation for both ordered *o1*-LNMO and *o2*-LNMO, we focused on extracting information on the extent of the ordering. For both datasets, ordering spots along the [110] direction is chosen, as they are commonly visible throughout the particles (marked with red dots in Fig. 2b-f). The selection of suitable spots, crucial for intensity analysis, is detailed in Fig. S12.

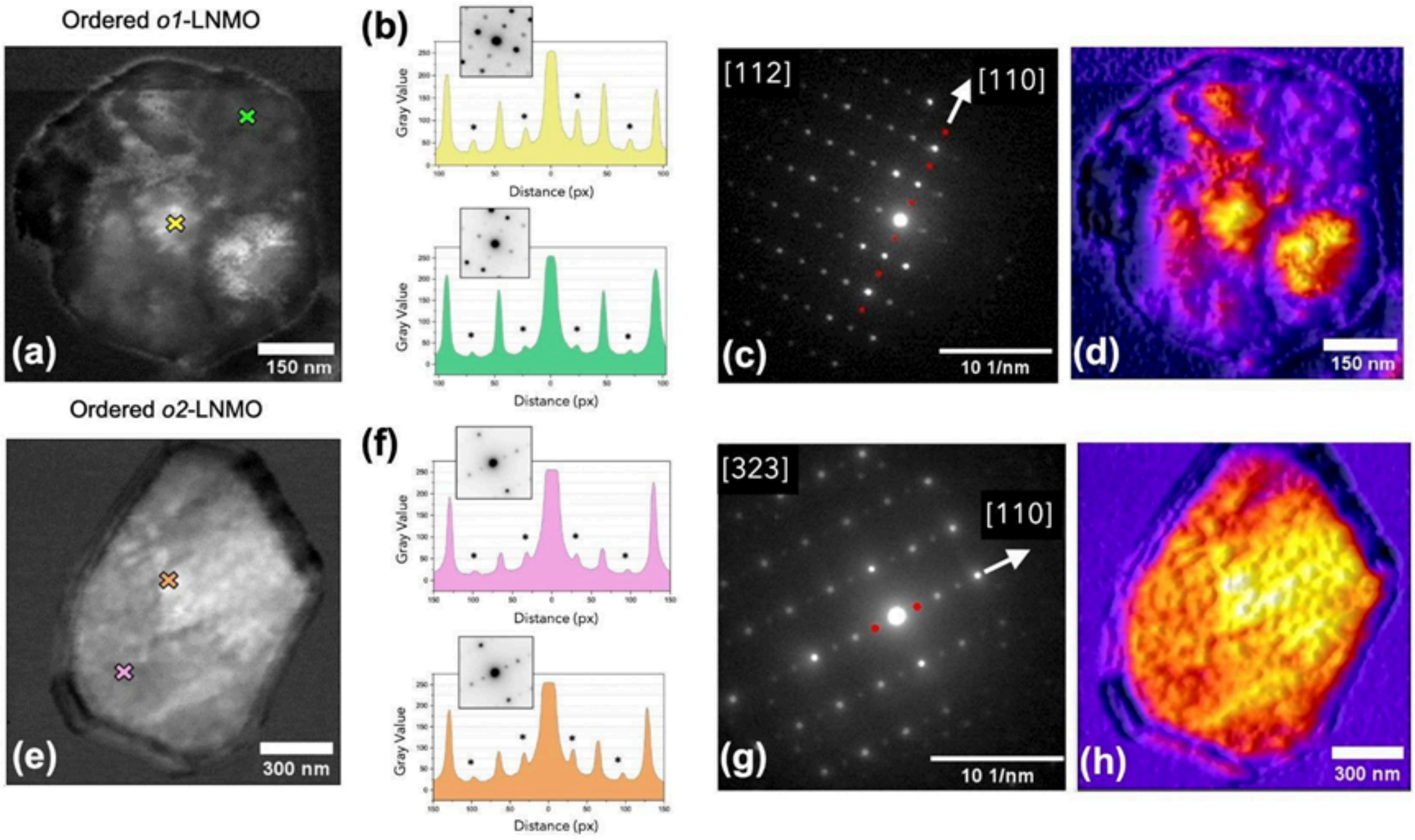

**Figure 2. Ordering intensity maps for LNMO indexed as $P4_332$ phase**. Process, including the selection of diffraction peaks for structural metal ordering mapping. **(a,e)** Virtual dark field images obtained from selected diffraction spots. Brighter zones indicate high intensity, while dark zones signify their absence. Two points are chosen on each particle, they correspond to the observed highest and lowest integrated ordering spots' intensities. The ordered *o1*-LNMO on the top (a), and the ordered *o2*-LNMO on the bottom (e). **(b,f)** Integration along the [110] direction on selected electron diffraction patterns given in (a,e). The ordering spots are marked with asterisks. The difference in their observed intensity between the integration of line profiles for each particle highlights the distribution of ordering degrees across the particle. **(c,g)** Selected diffraction spots (in red) for ordering intensity search by virtual dark field. The [110] direction is chosen on both zone axes because of its visibility around the particle. Selection details are given in Fig. S11. **(d,h)** Ordering intensity map represented with a colored scheme based on gray values. Multi-colored zones on particles indicate a heterogeneous ordering distribution with a spatial resolution of 10 nm.

The selected spots are then used as a virtual aperture, where a virtual dark field image is obtained by averaging the intensity of the collected diffraction pattern (Fig. 2c-g). Obtained virtual dark maps were then color-coded based on their gray values, as shown in Fig. 2d-h. Here, a white pixel corresponds to intensity saturation at 255 in grayscale, indicating a strong presence of ordering spots. On the other hand, a black pixel represents an intensity value of 0, indicating the absence of the searched ordering spots in that region. This provides a visual indication of the degree of ordering across the material, with brighter regions presenting a higher density of ordering spots.

In the case of the *o1*-LNMO, a heterogeneously colored intensity map is observed in Fig. 2d. The center of the particle exhibits a high concentration of intensity, as shown in the associated electron diffraction pattern in Fig. 2c. However, as we move toward the outer parts of the particle, the ordering spot intensity decreases, as in the example given in Fig. 2b (bottom). This observation indicates a variation in the degree of ordering within the particle, with higher intensity concentrated in the center and lower intensity towards the edges.

On the other hand, for the *o2*-LNMO, an overall high intensity is detected in the obtained intensity map shown in Fig. 2h. Similar to the *o1*-LNMO, the center of the particle exhibits the highest ordering spots' intensity, as shown in the associated diffraction pattern in Fig. 2g. However, in this case, when the outer parts of the particle are analyzed, there is a preservation of the ordering spots' intensity even in the lowest intensity zone (Fig. 2h).

**Automated quantification of ordering**

Two key limitations inherent to pattern-matching algorithms can still result in information loss during software-based phase mapping, even under ideal experimental conditions, and strongly impact the assessment of the degree of ordering in LNMO. First, a large matching radius may overlook weak extra spots due to Ewald sphere curvature, leading to incorrect phase identification (i.e., misattribution as a disordered domain). Second, intensity thresholds based on ideal models fail to account for experimental complexities like overlapping lattices and dynamical effects. These factors cause unexpected intensity variations that the software may misinterpret, especially in regions with weak signals. To overcome these limitations, we developed a data science strategy with the ePattern software[33]. This tool can individually detect Bragg spots both in position and in intensity, enabling a precise separation between principal and extra spots.

To quantify the degree of ordering (order parameter) at the particle level using the 4D-STEM ACOM dataset, we defined an order parameter, α, as the ratio between the mean intensity of the ordering-specific diffraction spots and that of the principal diffraction spots, which are common to both ordered and disordered lattices. To ensure comparability across measurements, all values were normalized with respect to a simulated diffraction pattern of a perfectly ordered reference structure, for which the maximum observable parameter was 0.8. This normalization was introduced through a scaling factor $\beta$ = (1 / 0.8) = 1.25, leading to following experission:

$$\alpha = \beta \frac{\langle I_e \rangle}{\langle I_p \rangle} \tag{1}$$

where $\alpha$ is the order parameter, $I_e$: is the ordering spot intensity and $I_p$: is the principle spot intensity.

For a quantitative extraction of the intensities, the initial step involves processing diffraction pattern data using ePattern[33], which extracts the positions and intensities of diffraction spots from each diffraction pattern, allowing the reconstruction of the diffraction patterns without noise. All diffraction patterns were recentered by aligning their central spot with the chosen reference center. We assess the number of principal and ordering spots, determine their intensities, and compute the order parameter according to eq. (1). The workflow is illustrated in Fig. S13. We have developed a python program (ePattern_Ordering) to facilitate spot selection within diffraction patterns using the output of the ePattern (a csv file containing the positions and intensities of each diffraction spot in our dataset). By extracting the coordinates of both principal and ordering spots after alignment of the central spot, we define circular regions of a fixed radius around these coordinates. This allows us to access to the intensity of each spot, then to classify a spot as "ordering" or "principal". We use a coordinates criterion, as the positions of the ordering and principal spots are based on the pattern calculated from the CIF file (Fig. S14). Fig. 3a illustrates the experimental setup for 4D-STEM, along with the process of selecting diffraction spots. This selection enables the computation of the order parameter, leading to the structural mapping of transition metal ordering. Fig. 3b shows the intensity distributions of the ordering spots, principal spots, and central spots for the *o1*-LNMO particle (orange shades) and the *o2*-LNMO particle (green shades).

The ordering spots in *o2*-LNMO exhibit higher intensities than those in *o1*-LNMO, whereas the principal spots show the opposite trend. This suggests a higher degree of ordering in *o2*-LNMO compared to *o1*-LNMO in agreement with previously discussed software-based results. The quantifications of the order parameter are shown in the maps in Fig. 3c and 3d, where different colors represent varying degrees of ordering within the LNMO particles. All order parameter values were normalized with respect to a simulated diffraction pattern of a perfectly ordered reference structure, for which the maximum observable order parameter was 0.80 (Fig S14). Notably, no zones without ordering were detected (blue regions indicate values higher than 0.0001). The in red zones indicate zones of uncertainty where elevated noise levels hinder the reliable detection of low-intensity ordering spots against the background.

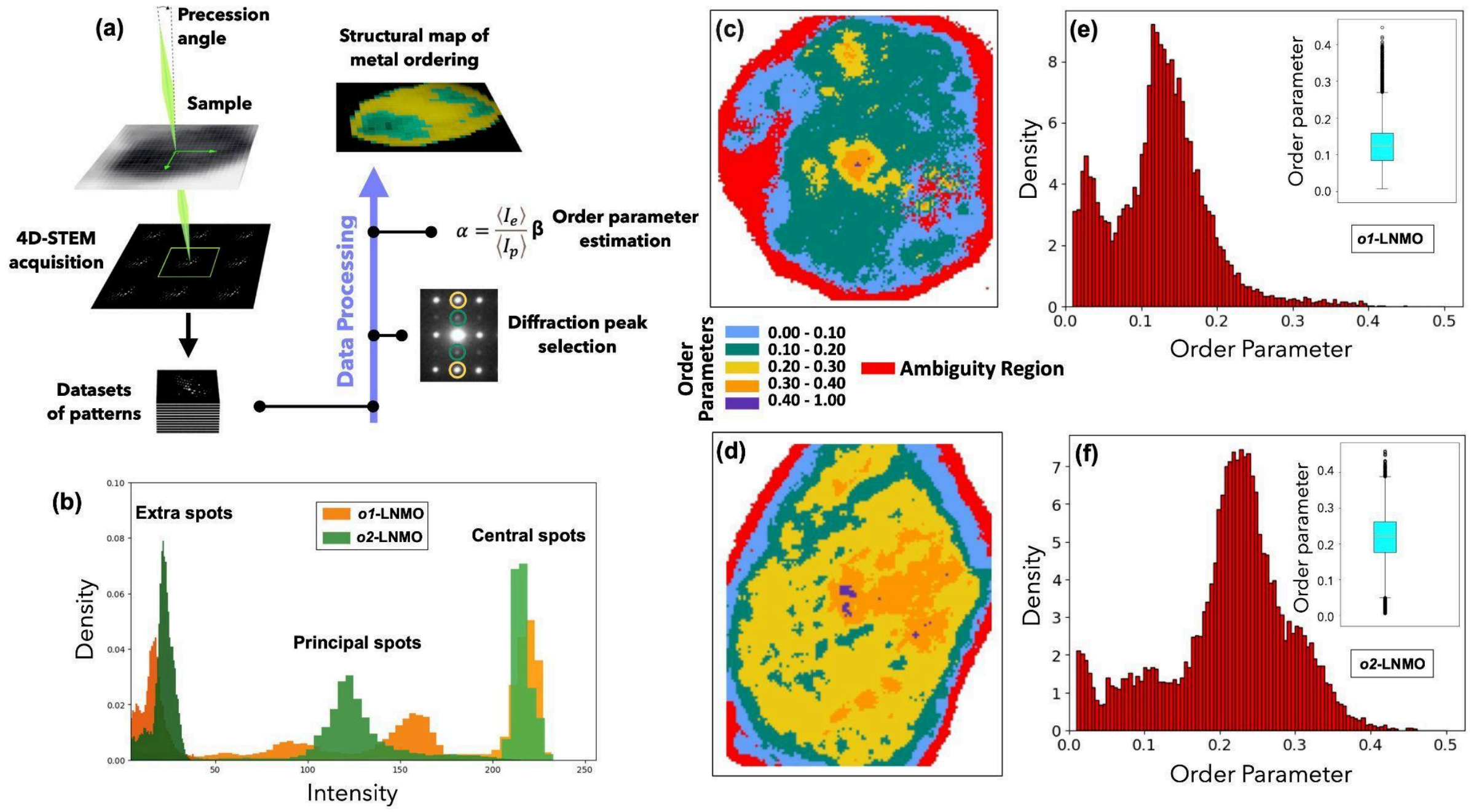

**Figure 3. Detailed analysis of order parameter maps for two LNMO samples. (a)** Schematic of the 4D-STEM acquisition process, including the selection of diffraction peaks for structural order mapping. **(b)** The intensity distribution for *o1*-LNMO is represented in orange shades, while *o2*-LNMO is depicted in green shades. Three distinct distributions are observed, one for the extra spots, one for the principal spots, and one for the central spots. **(c,d)** The corresponding distribution of order parameters for *o1*-LNMO and *o2*-LNMO particles, respectively. **(e,f)** Examination of the order parameter distribution (considering only values above 0.0001) provides further insights into the quantification values and the box plot highlights the order parameter outliers.

Fig. 3e and 3f display histograms of the order parameter distribution for *o1*-LNMO and *o2*-LNMO, respectively. In *o1*-LNMO, values range from approximately 0.0076 to 0.44, whereas *o2*-LNMO exhibits a broader distribution, extending from 0.0059 to ~0.46. The boxplot for *o1*-LNMO shows more outliers starting from approximately 0.26 and extending to higher values, with no outliers detected at lower values. The *o2*-LNMO boxplot exhibits outliers primarily at lower values, approximately below 0.05, and fewer outliers at higher values. In terms of data distribution, the box plot indicates that for *o1*-LNMO, the first 25% of values range from 0.0076 to approximately 0.09, while the middle 50% of the values lie between 0.09 and 0.15. For *o2*-LNMO, the lower 25% of values range from 0.0059 to around 0.17, and the central 50% spans from approximately 0.17 to 0.25. Overall, *o2*-LNMO tends to show a higher degree of ordering, as evidenced by the greater prevalence of orange and yellow regions in Fig. 3c and 3d, corresponding to order parameter values between 0.20 and 0.40. Notably, localized regions within *o1*-LNMO also exhibit high ordering, with values approaching 0.44, highlighting spatial heterogeneity in the transition metal ordering.

To enhance the reliability of our quantification method, we investigated the relationship between local thickness (Fig. S15) and zone axis misalignment (Fig. S16) with the calculated order parameter values. A comparison between the intensity of the central diffraction spot and the order parameter at each pixel revealed no direct correlation between the thickness and order parameter. The deviations from the ideal zone axis were effectively highlighted by comparing the intensities of opposing diffraction spots (*e.g.* the ordering spots corresponding to $(\bar{4}40)$ and $(4\bar{4}0)$). The ordered *o1*-LNMO particle exhibits two regions that are clearly off-zone axis, while the *o2*-LNMO particle remains largely aligned along the zone axis. These off-zone-axis regions (Fig. S16e and S16j) coincide with the ambiguous and low-order regions identified in the initial quantification maps (Fig. 3c and 3d). Notably, *o1*-LNMO also displays two distinct principal spot intensities in the distribution shown in Fig. 3b, which is attributed to the presence of a off-zone-axis area (Fig. S16c). This of-zone-axis area was previously identified as one of the highest-intensity regions in the virtual aperture selection for ordering spots (Fig. 2d). These observations highlight both the accuracy of our quantification method and the importance of validating its reliability.

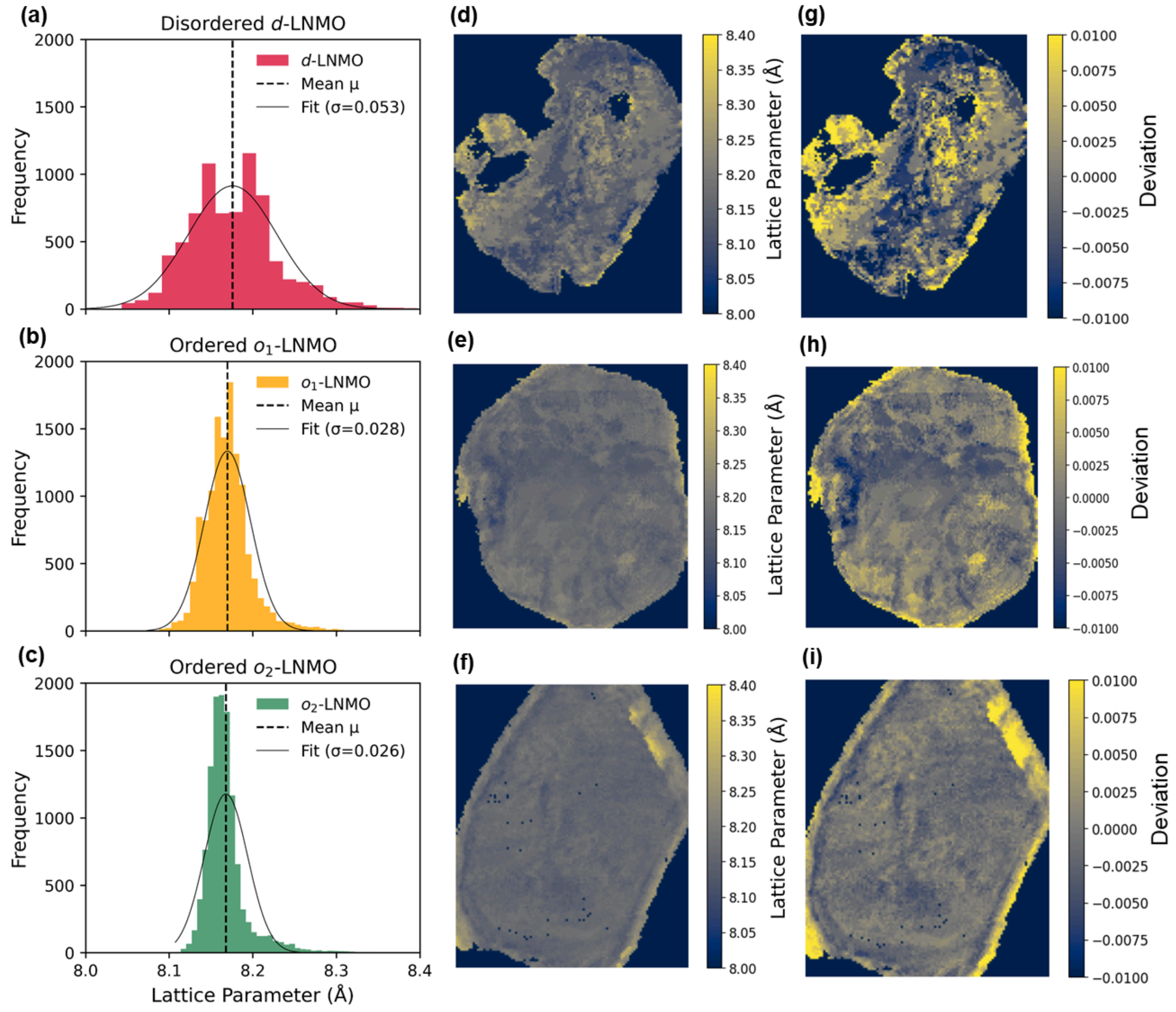

**Figure 4. Lattice parameter map calculated from 4D-STEM data. (a,b,c)** Lattice parameter distributions for *d*-LNMO (red), *o1*-LNMO (orange), and *o2*-LNMO (green). Dashed lines indicate the mean values obtained from X-ray powder diffraction, which were used to calibrate the average lattice parameters extracted from electron

diffraction data. Black curves show Gaussian fits with the extracted standard deviation (σ) used as a measure of distribution width. **(d,e,f)** Lattice parameter distribution maps for *d*-LNMO, *o1*-LNMO, and *o2*-LNMO, respectively. **(g,h,i)** Deviations of the lattice parameters from their average value for the corresponding particles.

Differences in lattice parameters serve as a reliable indicator of the composition in spinel LNMO. As discussed in the “Description of the Global Structure” section, the lattice parameter of disordered LNMO is typically larger than that of the ordered phase. Here, we calculated local lattice parameters with a spatial resolution of approximately 10 nm using 4D-STEM datasets, based on the diffraction spots highlighted in Fig. S17. Details of the pattern calibration procedure used to obtain absolute distances are provided in the Supplementary Information. Due to dynamical scattering and calibration uncertainties in camera length and detector geometry, lattice parameters obtained from electron diffraction can deviate in absolute value from those obtained from X-rays. Therefore, the electron diffraction values were calibrated against the average lattice parameters obtained from X-ray diffraction: 8.1759(2) Å for *d*-LNMO, 8.1697(6) Å for *o1*-LNMO, and 8.1679(4) Å for *o2*-LNMO.

Figure 4 presents the lattice parameter maps and corresponding distributions, illustrating the spatial variations for *d*-LNMO, *o1*-LNMO, and *o2*-LNMO. In the disordered particle, significant local variations in lattice parameter are evident, which may reflect compositional inhomogeneities such as variations in $Mn^{3+}$ content or local defects. By contrast, the *o2*-LNMO sample exhibits an exceptionally homogenous lattice parameter distribution, with a coefficient of variation (CV = $\sigma/\mu$) of 0.0032 compared to 0.0065 for *d*-LNMO. The largest deviations are observed at the particle edges, where high noise and/or low diffraction spot intensities were noted previously. Interestingly, although *o1*-LNMO is both globally and locally ordered and has an average lattice parameter close to that of *o2*-LNMO, its CV (0.0034) is higher, and its local lattice parameter distribution resembles that of the disordered *d*-LNMO particle. Similar to the order parameter analysis, the lattice parameter measurements therefore reveal considerable spatial heterogeneity in *o1*-LNMO.

## Discussion and conclusion

In this study, we have thoroughly investigated the spatial distribution of transition metal ordering in spinel $LiNi_{0.5}Mn_{1.5}O_4$ (LNMO) using advanced 4D-STEM combined with a data science-driven analytical pipeline. Starting from a globally disordered LNMO sample *d*-LNMO, two annealing conditions produced structurally ordered samples *o1*-LNMO and *o2*-LNMO, both indexed in the $P4_332$ space group by neutron diffraction. However, our 4D-STEM analysis reveals that despite their long-range crystallographic similarity, these two samples exhibit markedly different local ordering behaviors at the nanoscale. By integrating diffraction datasets with robust pattern matching and spot-intensity analysis, we present the first spatial mapping of ordering heterogeneity with a direct quantification of transition metal ordering within individual LNMO particles.

Spatially resolved phase and order parameter mapping with 10 nm resolution confirmed the presence of transition metal ordering across the particle thickness in both ordered samples. However, significant differences emerged. The *o2*-LNMO sample, annealed longer and under continuous oxygen flow, exhibited a consistently higher and more homogeneous order parameter. In contrast, *o1*-LNMO, annealed for a shorter time under ambient air, showed lower overall ordering and pronounced spatial heterogeneity. Notably, its local lattice parameter distribution closely resembled that of the disordered *d*-LNMO, suggesting incomplete ordering. It is important to note that these trends were consistently observed across multiple particles per sample, confirming the statistical robustness of our findings.

These observations challenge previous structural models that propose uniform coherent ordering in LNMO crystallites. Our data suggest that the global ordering behavior is governed at the nanometric scale, aligning with findings by Tertov et al.[35], who observed a distribution of nanosized ordered domains separated by disordered regions using aberration-corrected TEM. However, within our nanometric resolution limit, we did not observe isolated disordered domains surrounding larger ordered domains, as reported in their study. This difference likely arises from specimen thickness: their analysis was performed on relatively thin lamellae (~50 nm), comparable to the size of individual domains, which enabled direct imaging of ordered and disordered regions. By contrast, our 4D-STEM measurements were conducted on significantly thicker samples (~180-200 nm), where overlapping contributions from multiple domains are unavoidably averaged through the sample thickness.

If complete ordering extended through the particle thickness, the detected spot intensities would remain uniform. In the case of the ordered samples, however, especially in *o1*-LNMO, we detect significant variations in the ordering parameter, which we attribute to the coexistence of ordered and disordered domains along the thickness of the particle.

The contrasting behaviors of *o1*- and *o2*-LNMO indicate that oxygen partial pressure is a critical factor: higher oxygen availability during annealing appears to suppress defect formation and stabilize the ideal, ordered phase. Interestingly, no secondary phases or rocksalt-type defects were detected in either *o1*-LNMO or *d*-LNMO. Quantitative analysis shows that $Mn^{3+}$ content decreases from 4.9 % in disordered d-LNMO to ~3.2 % in *o1*-LNMO and further to ~1.1 % in highly ordered *o2*-LNMO. This compositional trend is consistent with the lattice parameter distributions calculated from 4D-STEM analysis: *d*-LNMO shows significant local variation with higher values, *o1*-LNMO retains a heterogeneous distribution resembling the disordered state, while *o2*-LNMO exhibits a highly homogenous distribution. Together, these results indicate that the transformation from *d*-LNMO to *o1*-LNMO can be attributed to partial ordering of Ni/Mn on the 4d/12d sites, accompanied by a moderate reduction of $Mn^{3+}$ content, whereas the homogeneity in ordered domains as in the *o2*-LNMO sample requires extended annealing under oxygen flow.

A novelty of this work is integrating a data science-based methodology to overcome limitations in 4D-STEM phase mapping, such as intensity thresholding and spot matching radius issues. Using the ePattern software, we developed a robust framework to detect individual Bragg spots in position and intensity. This was coupled with a custom Python routine (*ePattern_Ordering*) to compute a local order parameter (α) based on the intensity ratio of ordering-specific to principal diffraction spots. To ensure accurate quantification, we addressed key experimental limitations, namely, sample thickness and crystal orientation, both of which can obscure ordering detection through dynamical scattering or off-zone-axis effects. These challenges were mitigated in our case by performing precession-assisted 4D-STEM on thin, platelet-shaped LNMO particles and a systematic analysis by considering each technique's limitations.

Altogether, this work offers new insights into the nanoscale structural complexity of LNMO and establishes a reproducible and transferable methodology for mapping local orderings. Our findings provide a new framework to understand the interplay between synthesis conditions, local structure, and electrochemical performance, which is essential for the rational design of battery materials.

## METHODS

### Material Preparation

LNMO materials were synthesized using molten salt synthesis as described in detail in our previous work[36]. The $Ni(NO_3)_2{\cdot}6H_2O$ and $Mn(NO_3)_2{\cdot}4H_2O$ precursors were mixed in stoichiometric amounts (1:3) with an excess amount of LiCl salt with a ratio *number of moles of LiCl over the number of moles of transition metals* of 35. The powder mixture was calcinated at 750°C for 4h in a tubular furnace under an air atmosphere. At the end of synthesis, the residual salt was eliminated by washing with water and then ethanol. The obtained sample was named the *disordered d*-LNMO according to the description of its crystal structure in the $Fd\bar{3}m$ space group. To order the transition metals, this material *d*-LNMO was re-annealed at 700°C for 4h (two times to homogenize the material) under air atmosphere or for 24 h under continuous $O_2$ flow, they were named *ordered o1*-LNMO and *ordered o2*-LNMO, respectively. All samples share a common platelet morphology characterized by an average size of 1 µm and a thickness of 180 nm.

## 4D-STEM and TEM techniques

Two different microscopes were used in this study. The electron diffraction patterns given on Fig. S4 in supplementary information were collected on a transmission electron microscope JEOL JEM-2100 (LaB6) at 200kV, using the selected area technique (SAED) with a Gatan Orius 200D camera. 4D-STEM ACOM studies were conducted on a FEI Tecnai F20 TEM at 200 kV equipped with NanoMEGAS Astar system and Gatan Oneview CMOS camera. For the 4D-STEM study, the spot size was set to 5, gun lens to 3 and a 10 µm condenser aperture (C2) was used (semi-convergent angle of 0.4 mrad). To reduce dynamical effects, the electron beam was processed at an angle of 0.8°. A step size of 10 nm (the width of one pixel) was set for all scans, defining the resolution of the obtained maps.

## Order Degree Quantification

The data science workflow for 4D-STEM ACOM data, as illustrated in the schematic in Fig. S13, is structured into two main phases: ePattern_Registration and ePattern_Ordering. This approach enables high-precision, quantitative analysis of crystallographic ordering at the nanoscale by leveraging automated peak detection, spatial alignment, and custom intensity-based metrics.

The first phase, ePattern_Registration, begins with the identification and quantification of diffraction spots. In this step, the ePattern software processes each diffraction pattern collected across the 4D-STEM scan, determining the exact position and intensity of every Bragg spot. This step is essential for generating a clean, standardized dataset that captures both the structural information and spatial variation present in the sample. Following this, a pattern alignment procedure is applied to ensure consistency across the dataset. All patterns are aligned based on the coordinates of the central (direct) beam spot, effectively centering the dataset and correcting for any minor shifts or drift during acquisition. This spatial normalization ensures that all subsequent comparisons between spots are meaningful and accurate.

The second phase, ePattern_Ordering, focuses on the evaluation and mapping of the degree of ordering using a calculated metric called the order parameter. This begins with the definition of the selection zone, where specific Bragg spots are selected for analysis. Using custom-developed algorithms, the user or software can define circular or rectangular regions that isolate the relevant ordering-specific spots (e.g., those present in the ordered phase but forbidden in the disordered one) and the principal spots common to both phases. Once the spot coordinates are established, the order parameter ($\alpha$) is calculated at each scan point using the intensity ratio between these two groups of reflections. Specifically, $\alpha$ is defined as the mean intensity of the extra (ordering-specific) spots divided by the mean intensity of the principal spots. This parameter serves as a quantitative proxy for the degree of local cation ordering within the sample. Finally, the computed $\alpha$ values are used to reconstruct spatial maps that visualize the ordering distribution across the entire particle with nanometric resolution.

## X-Ray Diffraction

The powder X-Ray Diffraction measurements were performed on a PANalytical Empyrean Diffractometer with Cu Kα radiation ($\lambda_{K\alpha1}$=1.54056 Å and $\lambda_{K\alpha2}$=1.54439 Å) in Bragg-Brentano configuration. The diffraction patterns were collected in a 2ϑ range of 15-120°(2ϑ) with a scan step of 0.008° for a total acquisition of 20 hours. Structural parameters were calculated through Le Bail refinements using the FullProf program[37]. The spinel domain sizes were calculated from the Scherrer's equation[38] after removing the instrumental contribution.

## Powder Neutron Diffraction

The room temperature powder neutron diffraction was performed on the D2B high resolution diffractometer of the Institut Laue-Langevin in Grenoble, France in transmission mode and using a wavelength of 1.594 Å. The data were evaluated via full pattern matching using the FullProf Program (Le Bail method). The ordered domain sizes were calculated using a selective broadening model (size-model=-14) using the Williamson-Hall method, a description of the peak profiles by the Thomson-Cox-Hastings Pseudo-Voigt function and an instrumental resolution file.

## Electrochemistry

The positive electrode was prepared by mixing 80 wt.% of LNMO powder, 10 wt.% of carbon black conductive additive (Alfa Aesar, 99.9 %), and 10 wt.% of Poly(vinylidene fluoride) (PVDF, Sigma-Aldrich) binder in 1-methyl-2-pyrrolidone as solvent (NMP, Sigma Aldrich, 99.5%). CR2032-type coin cells were assembled in an Ar-filled glovebox using Li metal as negative electrode, Whatman® separator and 1M $LiPF_6$ EC/DMC electrolyte (1:1 w/w ethylene carbonate (E.C.) / dimethyl carbonate (DMC)). The electrochemical performance of LNMO materials was assessed using the galvanostatic charge-discharge method. All the electrochemical experiments were conducted at 25°C.

## Acknowledgments

The authors thank the Région Nouvelle Aquitaine and the ANR French National Research Agency (DESTINa-ion_Operando project ANR-19-CE42-0014-02 and Labex STORE-EX project ANR-10-LABX-76-01) for their financial support. They also thank the ANR for the funding of G.O. and C.Z.'s PhD and postdoctoral fellowships, respectively. The authors thank Ilia Tertov from ICMCB (Pessac, France) for fruitful discussions and technical support on neutron diffraction. The authors also thank Nicolas Folastre from LRCS (Amiens, France) for fruitful discussions and technical support on 4D-STEM.

## Author Contributions

G.O. synthesized the analyzed materials, carried out the X-ray and electron diffraction experiments, and the electrochemical measurements, and interpreted the data across all techniques. F.A. quantified the local order parameter and calculated lattice parameter distributions from 4D-STEM data, developed dedicated 4D-STEM analysis codes for these analyses, and interpreted the data. G.O. and C.Z. performed the 4D-STEM data acquisitions with A.D. and M.V.. J.C. performed the non-negative matrix factorization clustering analysis. E.S. was responsible for the neutron diffraction data acquisition. G.O. wrote the first draft of the manuscript. J.O., L.C., F.W., and A.D. supervised the project and interpreted the results. All authors took part in the scientific discussion and contributed to the writing and revision of the manuscript.

# Supplementary Information

The disordered *d*-LNMO is obtained by a synthesis in molten salt as described in our previous work[1]. To order the transition metals, *i.e.* Ni and Mn, in the spinel structure, two different heat treatments were applied to the pristine *d*-LNMO sample. The first *ordered o1*-LNMO sample is obtained after two consecutive 4 hours annealings at 700°C in air atmosphere with intermediate grinding, while the *ordered o2*-LNMO sample is prepared after two consecutive 24 hours annealings at 700°C under continuous oxygen flow.

The lattice parameters calculated from X-Ray diffraction patterns (**Figure S1a**) are 8.1759(2) Å, 8.1697(6) Å and 8.1679(4) Å for disordered *(d-)*, ordered *(o1-)* and ordered *(o2-)* LNMO respectively. These values are in agreement with the ones reported in the literature.[2,3] The slight variations in lattice parameters among the samples are generally attributed to differences in $Mn^{3+}$ content and thus in Ni/Mn composition in the spinel phase, and thus to differences in the nature and amount of impurities such as lithiated rock salt or layered phases to compensate. Nevertheless, despite these different lattice parameters and similar precursors' mixtures used for their syntheses, no impurity phase for LNMO[4] was detected for none of the three samples. It suggests in our case the presence of defects with crystalline domains' size small enough not to be observed by XRD. The crystalline domains' sizes of the spinel structure were calculated using Scherrer's equation and no significant change was observed between samples (63, 60 and 55 nm *d*-, *o1*- and *o2*-LNMO*,* respectively).

However, the very different neutron scattering factors of Ni and Mn atoms ($b_{Mn}$= -3.73 fm and $b_{Ni}$= 10.3 fm)[5] (in magnitude and in sign) allow us to observe in the neutron diffraction patterns the additional diffraction peaks resulting from the ordering of these atoms (**Figure S1b**). Therefore, the two annealed samples (*o1*-LNMO and *o2*-LNMO) can be indexed within the $P4_332$ space group, as an ordering of the transition metals happens during the annealing of *d*-LNMO that is fully disordered, as revealed by the absence of the additional lines. The composition of the spinel phases was evaluated using Rietveld method, and the results are reported in **Figure S2** and **Table S1** hereafter. The spinel phase in *o2*-LNMO sample presents a Ni/Mn ratio of 0.49/0.51, close to the stoichiometric one of 0.5/1.5, while the *d*-LNMO and the *o1*-LNMO samples' spinel phases exhibit a slight over-stoichiometry in Mn with a Ni/Mn ratio of 0.46/1.53 and 0.48/1.52, respectively. Note that a significant difference in the full width at half maximum (FWHM) of the peaks characteristic of the ordering can be observed between *o1*- and *o2*-LNMO samples, as highlighted in the inset in **Figure S1b**. The coherent ordered domain sizes, calculated from the order-related peaks, were found to be 30 nm for *o1*-LNMO and 45 nm for *o2*-LNMO.

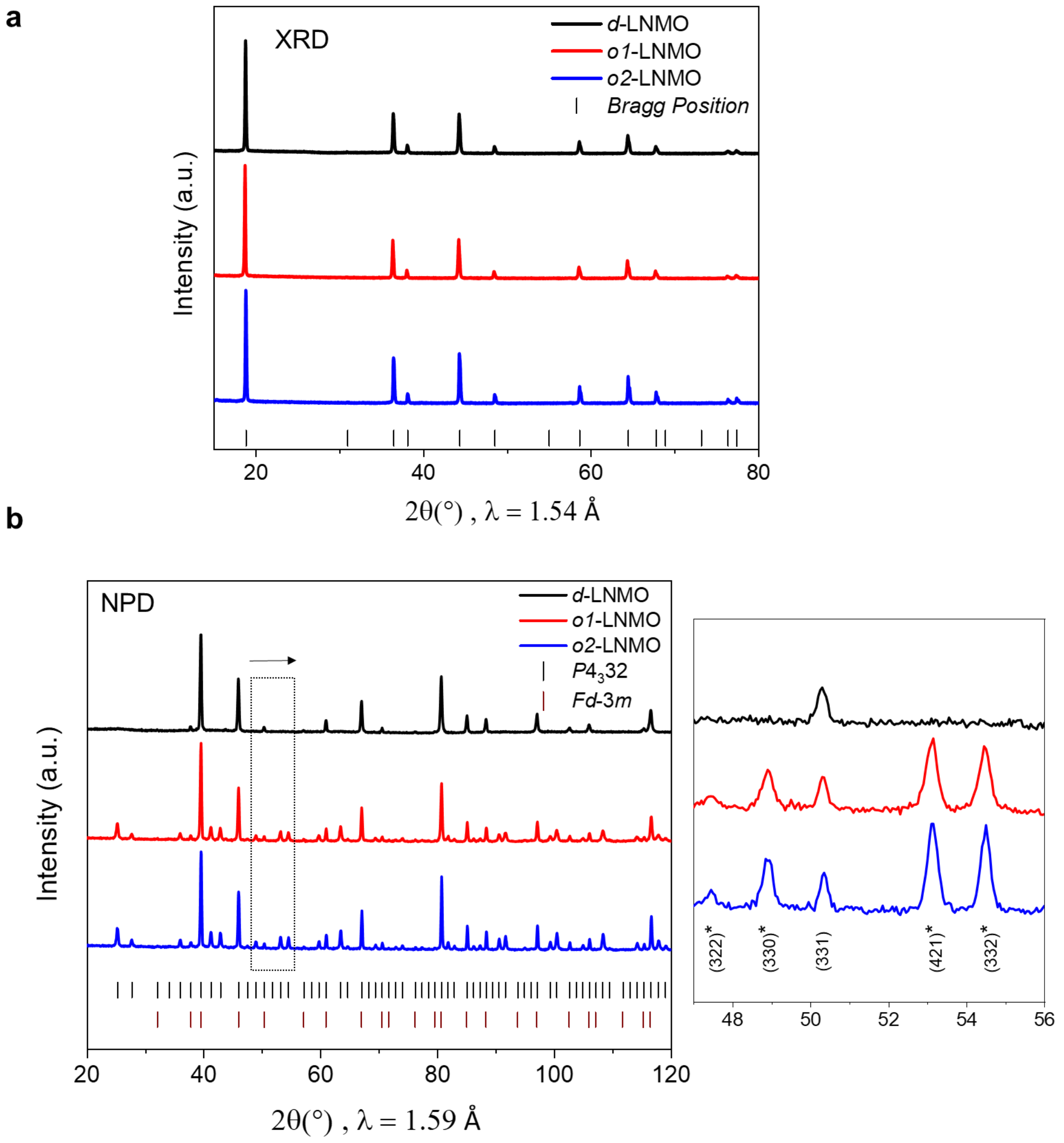

**Figure S1: Averaged information on the degree of ordering. a.** X-ray diffraction patterns of *disordered* *d*-LNMO (black), *ordered o1*-LNMO (red), and *ordered o2*-LNMO (blue). All patterns are indexed with cubic spinel $LiNi_{0.5}Mn_{1.5}O_4$ phase (JCPDS card no. 80-2162). The Bragg positions are given for *Fd*$\bar{3}$*m* space group. **b.** Neutron powder diffraction patterns of these LNMO samples. According to the Bragg positions, that of *d*-LNMO is indexed in the *Fd*$\bar{3}$*m* space group, while those of *o1*- and *o2*-LNMO are indexed in the $P4_332$ space group. All patterns are normalized to the highest-intensity reflection at 39.5°. A zoom-in view (right) shows differences in the intensity and width of the ordering peaks corresponding to the reflections highlighted by an asterisk.

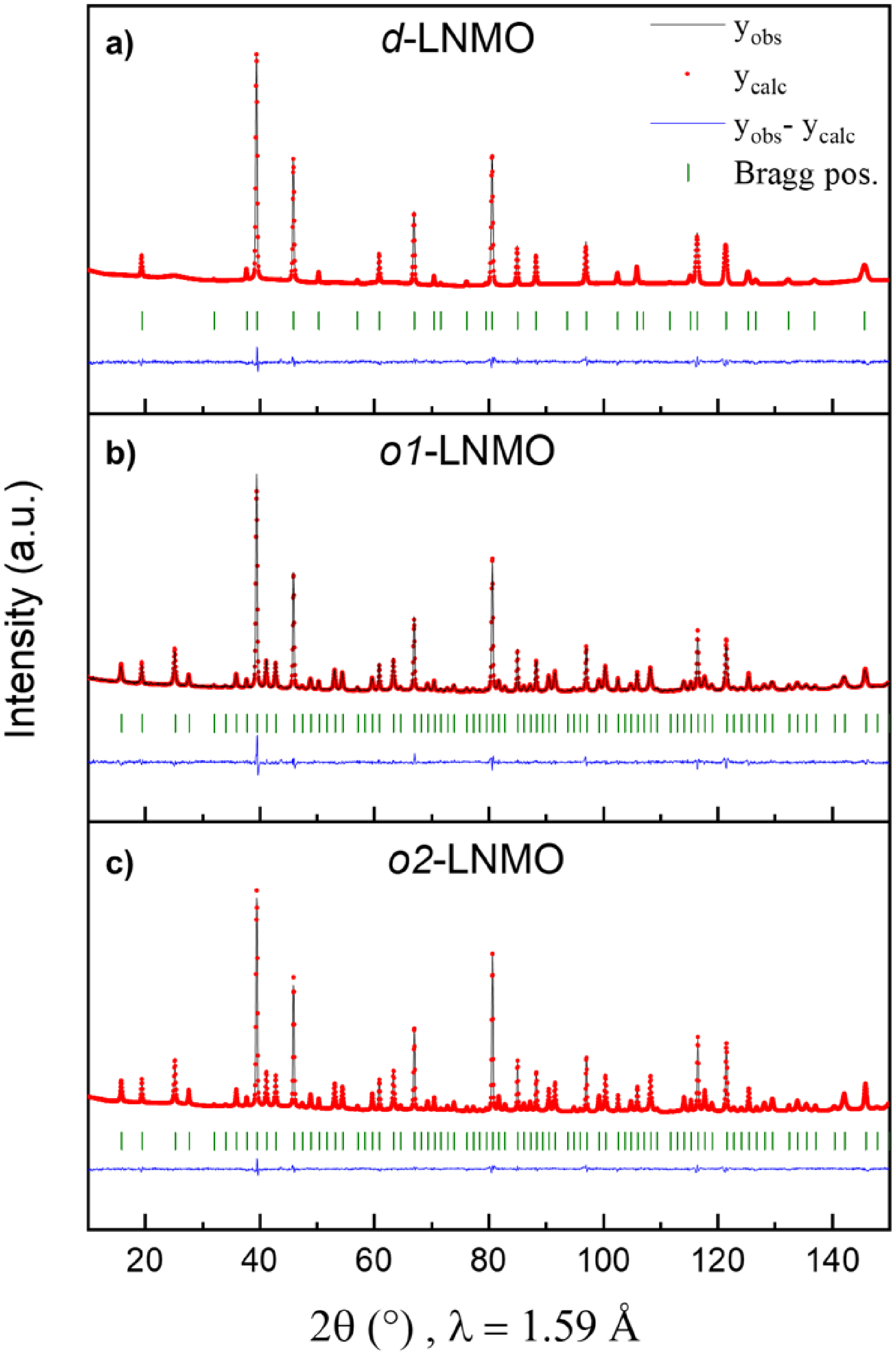

**Figure S2: Rietveld fits of the neutron powder diffraction patterns.** **a** Refinement for the sample *d*-LNMO in $Fd\bar{3}m$ space group, **b** for the sample *o1*-LNMO and **c** for the sample *o2*-LNMO, both described in $P4_332$ space group.

| Sample | Atom | Wyckoff position | x | Y | z | $B_{iso}$, Å$^2$ | Occ. |
|---|---|---|---|---|---|---|---|
| ***d*-LNMO** | Li | 8*a* | 1/8 | 1/8 | 1/8 | 1.3(2) | 1 |
| ***Fd*$\bar{3}$m** | Mn | 16*d* | 1/2 | ½ | 1/2 | 0.6 | 0.769(2) |
| **(#227)** | Ni | 16*d* | 1/2 | ½ | 1/2 | 0.6 | 0.231(2) |
| | O | 32*d* | 0.2630(11) | 0.2630(11) | 0.2630(11) | 0.9(2) | 1 |
| $R_p$ = 26.9%, $R_{wp}$ = 16.9%, $R_B$ = 9.5%, $\chi^2$ = 1.95 | | | | | | | |
| | Li | 8*c* | 0.0032(9) | 0.0032(9) | 0.0032(9) | 0.9(7) | 1 |
| | Mn | 12*d* | 1/8 | 0.6248(3) | 1.1150(9) | 0.3(5) | 1 |
| ***o1*-LNMO** | Ni | 4*b* | 5/8 | 5/8 | 5/8 | 0.6(4) | 0.952(2) |
| ***P*$4_3$32** | Mn2 | 4*b* | 5/8 | 5/8 | 5/8 | 0.6(4) | 0.048(2) |
| **(#212)** | O1 | 8*c* | 0.3843(2) | 0.3843(2) | 0.3843(2) | 0.7(2) | 1 |
| | O2 | 24*e* | 0.1493(2) | -0.1423(2) | 0.1242(2) | 0.7(2) | 1 |
| $R_p$ = 16.7%, $R_{wp}$ = 15%, $R_B$ = 4.5%, $\chi^2$ = 2.21 | | | | | | | |
| | Li | 8*c* | 0.0035(9) | 0.0035(9) | 0.0035(9) | 0.8(2) | 1 |
| | Mn | 12*d* | 1/8 | -0.371(3) | 0.1161(9) | 0.2(6) | 1 |
| ***o2*-LNMO** | Ni | 4*b* | 5/8 | 5/8 | 5/8 | 0.4(4) | 0.984(2) |
| ***P*$4_3$32** | Mn2 | 4*b* | 5/8 | 5/8 | 5/8 | 0.4(4) | 0.016(2) |
| **(#212)** | O1 | 8*c* | 0.3843(2) | 0.3843(2) | 0.3843(2) | 0.5(2) | 1 |
| | O2 | 24*e* | 0.1502(2) | -0.1421(2) | 0.1245(2) | 0.5(2) | 1 |
| $R_p$ = 19.1%, $R_{wp}$ = 16.5%, $R_B$ = 7.97%, $\chi^2$ = 2.97 | | | | | | | |

**Table S1: Rietveld refinement results.** For the disordered *d*-LNMO sample (top), the spinel structure is described in *Fd*$\bar{3}$*m* S.G., the Mn and Ni occupancies being evaluated in the 16*d* octahedral site. Here, the atomic displacement parameters of transition metals could not be refined simultaneously, they were thus kept fixed. For the two ordered samples, *o1*-LNMO (middle) and *o2*-LNMO (bottom), whose spinel phase is described in *P*$4_3$32 S.G., the ratio between the Ni occupying the 4*b* octahedral site and Mn occupying the 12*d* octahedral site was evaluated. For all refinements, Li and O atoms occupancies were evaluated and found to tend to a full occupation of the corresponding site, they were thus kept fixed for the subsequent refinements.

The second charge profiles given in **Figure S3a** represent better the material's performance since during the initial charge, the formation of cathode electrolyte interphase (CEI) can alter the observed capacity for the samples. At the end of second charge, the obtained capacity is 146 $mAh.g^{-1}$ for *disordered d*-LNMO, while it decreases to 137 and 135 $mAh.g^{-1}$ for *ordered o1*- and *o2*-LNMO, respectively. This lower capacity is commonly observed when ordering of transition metals occurs in spinel LNMO.

First, all three samples undergo a solid solution reaction associated with the redox activity of $Mn^{3+}$ to $Mn^{4+}$ at approximately 4.1 V. This represents 4% of the total capacity for *disordered d*-LNMO while it decreases to only 1% for *ordered o2*-LNMO. Interestingly, *ordered o1*-LNMO displays two distinct shoulders at 3.96 and 4.2 V (shown by arrows in the inset of **Figure S3a**), constituting a total of 4% of its overall capacity. Despite this phenomenon is not common, it has been observed in previous research.[6,7] Although its origin remains unknown, it appears in Mn-rich ordered spinels. This discrepancy points to a significant difference in the composition, i.e. the Ni/Mn ratio in the spinel phase, as observed through NPD refinements, between our two ordered LNMO samples while also hinting possible defects within the sample *o1*-LNMO.

The samples also differ by their Ni redox activity where the *disordered* d-LNMO described in $Fd\bar{3}m$ space group undergoes initially a solid-solution reaction in the composition range $0.04 < x < 0.54$ in $Li_{1-x}Ni_{0.5-\alpha}Mn_{1.5+\alpha}O_4$, followed by a biphasic reaction from $Li_{0.46-x}Ni_{0.5-\alpha}Mn_{1.5+\alpha}O_4$ towards completely delithiathed $Ni_{0.5-\alpha}Mn_{1.5+\alpha}O_4$. Conversely, the two ordered samples described in $P4_332$ space group exhibit two plateaus, the signature of biphased reactions, associated with the redox activity of $Ni^{2+}$ to $Ni^{3+}$ followed by that of $Ni^{3+}$to $Ni^{4+}$. The solid-solution region in the disordered spinel LNMO phase is observed at a lower potential than the biphasic region in the ordered spinel LNMO phase (**Figures S3b-S3d**). This leads to a more significant voltage gap between $Ni^{2+}/Ni^{3+}$ and $Ni^{3+}/Ni^{4+}$ couples for the disordered phase. The voltage gap is 60 mV for *d*-LNMO while for both *o1*- and *o2*-LNMO, it decreases to 20 mV, indicating a higher degree of ordering in the structure. Even though there is no difference in voltage gap between the two ordered samples, there are differences in the redox mechanisms as revealed by the derivative curves: the highest intensity is observed for *o1*-LNMO, while the redox reactions occur at a slightly higher potential for *o2*-LNMO.

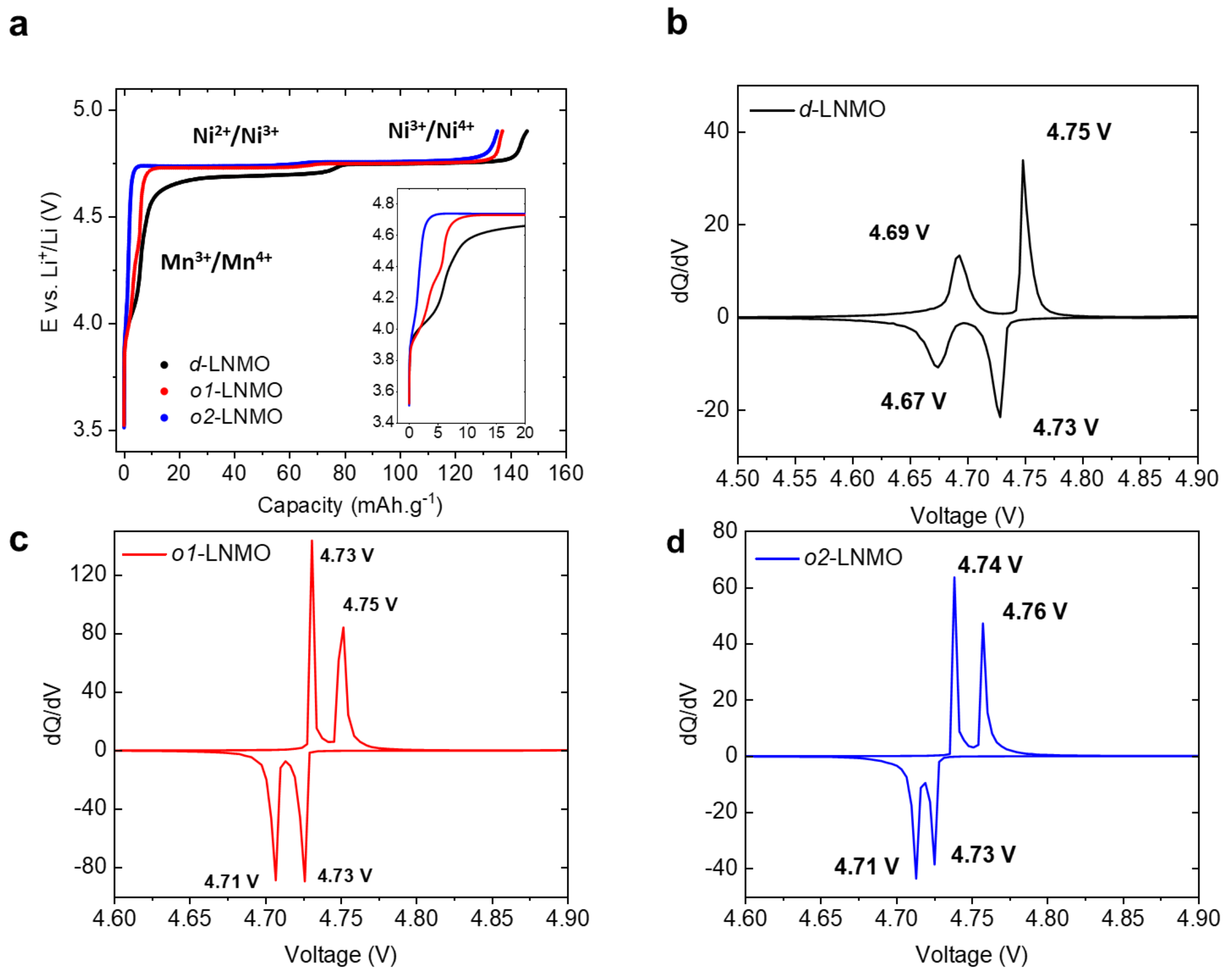

**Figure S3: Electrochemical performance comparison of the LNMO samples. a** Comparison of the 2nd charges obtained for the three LNMO samples at a C/10 cycling rate (i.e. 10 hours for the theoretical exchange of one $Li^+$ ion). A zoom-in view of Mn redox activity is given in the inset. **b, c** and **d** Corresponding dQ/dV curves of *d-, o1-*, and *o2*-LNMO in the voltage range of the Ni redox activity, respectively.

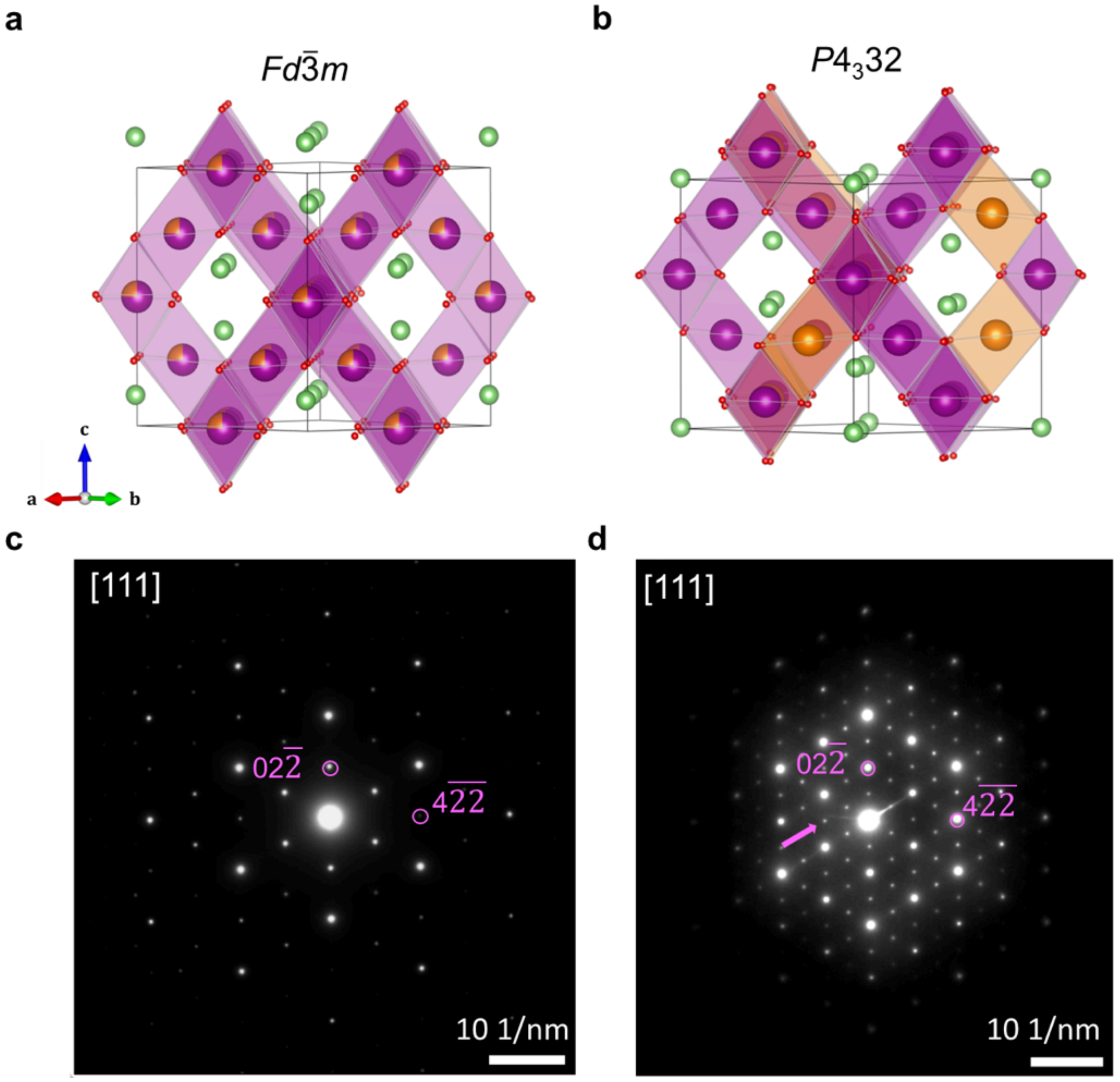

**Figure S4: Structural differences between disordered and ordered LNMO a-b** Crystal structures of disordered LNMO described in the $Fd\bar{3}m$ space group and ordered LNMO described in the $P4_332$ space group, respectively. Lithium atoms are represented by green balls, manganese by purple balls, nickel by orange balls, and oxygen by red balls. **c-d** Electron diffraction patterns obtained along the [111] zone axis and associated with disordered (c) and ordered (d) lattices. The pink arrow on (d) shows an example of an extra diffraction spot $(\bar{2}11)$ observed due to an ordering of the transition metals in a $P4_332$ arrangement.

## 4D-STEM ACOM parameters[8,9]

The acquired diffraction patterns were matched with simulated theoretical diffraction patterns (template banks), followed by the generation of phase and orientation maps using DiffGen2, Index2, and MapViewer2 within the NanoMEGAS software package. For accurate comparison of the data, the same pattern-matching parameters (threshold, softening loops, and gamma correction values) were applied on all samples. The quality and accuracy of the match are determined by two main parameters: correlation and reliability index representing the degree of matching between calculated templates and experimental data, respectively by comparison of the intensity and position of the diffraction spots and the comparison of the correlation index of the two for the best solution[8].

First, the theoretical electron diffraction patterns called templates are calculated within the software for both crystallographic structures, i.e. $Fd\bar{3}m$ (n°227) and $P4_332$ (n°212), under kinematical considerations. For one structure, around 2000 templates are generated with an approximate 1° of resolution. After the experimental data collection, each pattern is indexed using an algorithm based on the calculation of a correlation index denoted as Q. This latter depends on the position and intensity match between the observed and calculated Bragg spots. The experimental pattern's intensity function is represented by P(x,y), and each calculated template i is given by the function $T_i(x,y)$. This parameter is commonly used to differentiate the amorphous and crystalline zones, in our case the TEM grid (with Q <100) and the sample (Q<1000).

$$Q(i) = \frac{\sum_{j-1}^{m} P(x_j, y_j) T_i(x_j, y_j)}{\sqrt{\sum_{j-1}^{m} P^2(x_j, y_j)} \sqrt{\sum_{j-1}^{m} T_i^2(x_j, y_j)}}$$

Additionally, to evaluate the quality of the match, a reliability factor denoted as R is employed for both orientation and phase maps. This parameter compares the ratio between the correlation indexes $Q_1$ and $Q_2$ of the best matching solutions. The range of reliability is defined by the user in data treatment. Typically, values between 15 and 100 are considered confident, whereas a value below 15 indicates that multiple solutions have similar quality of matches.

$$R = 100(1 - \frac{Q_2}{Q_1})$$

**In our mappings, we maintained identical reliability ranges for each particle to ensure accurate comparison. Specifically, the orientation reliability was set between 15 to 70, while for phase mappings, this range was narrowed to 10 to 30 due to inherent similarity between two analyzed structures.**

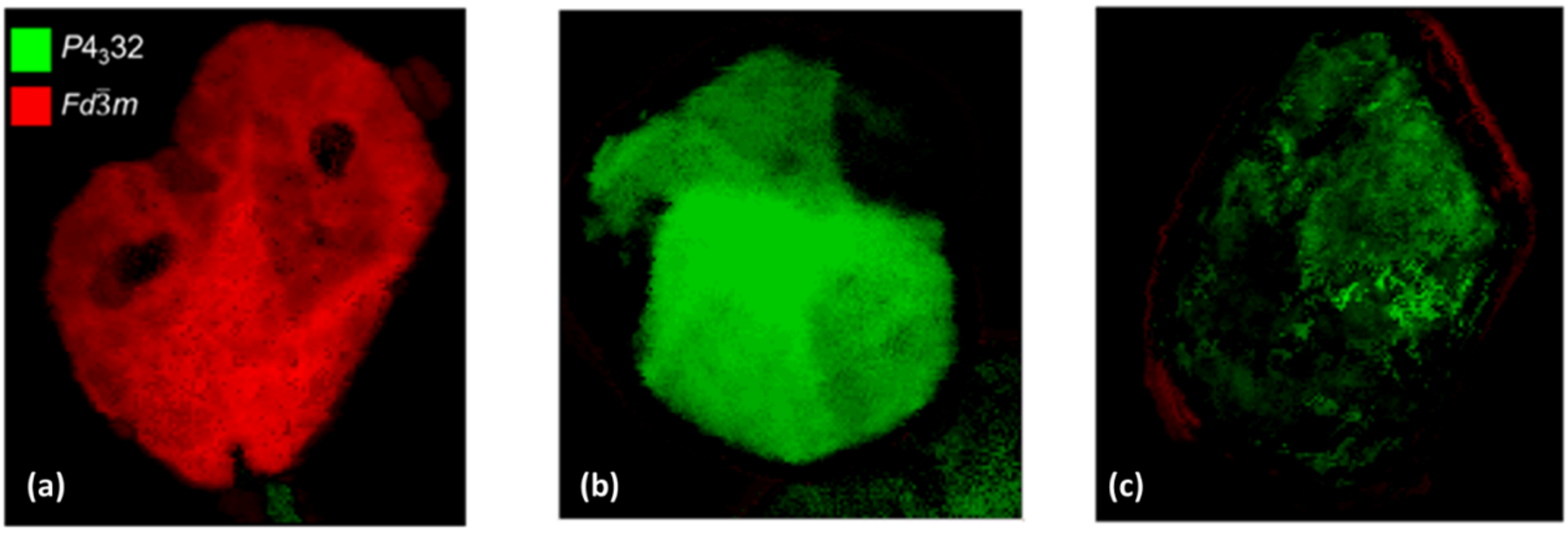

**Figure S5: (a-b-c)** Corresponding phase maps combined with phase reliability factor. All maps given are at the same scale as their associated TEM images given in the main text in Fig. 1b-f-i. Red zones are detected as corresponding to the $Fd\bar{3}m$ phase and green zones to the ordered $P4_332$ phase. Dark regions are detected as non-reliable zones within the chosen value (range 10-30 for all maps). The elimination of large zones by considering the reliability factor, and the reminiscence of zones indexed as being the $Fd\bar{3}m$ phase despite in ordered *o2*-LNMO map are further discussed in Fig. S8, S9, S10 and S11.

Fig. S5 shows the generated phase maps corresponding to the LNMO particles selected in Fig. 1(b-f-j), after comparison with the two theoretical diffraction patterns and applying an internal reliability factor in ACOM post-treatment software. All maps exhibit a dominant color scheme indicating the detection of one dominant phase in each case: the $Fd\bar{3}m$ phase (in red) for the disordered *d*-LNMO sample and the $P4_332$ phase (in green) for the two ordered *o1*-LNMO and *o2*-LNMO samples. The internal reliability factor quantifies the quality of the match by comparing the position and intensity of observed and expected diffraction spots. After considering this parameter, we observe the exclusion of multiple zones, meaning that not enough information is available on these zones to identify their structure without ambiguity. One reason can be for instance an overlap between particles and thus between their electron diffraction patterns that makes impossible for the software the accurate identification of the correct phase. This situation is for instance clearly illustrated in the case of *d*-LNMO, where smaller particles sitting underneath the large one can be distinguished by the contrast difference in the TEM image in Fig. 1b in the main text. The even more extended elimination of zones for *o1*- and *o2*-LNMO particles comes from a combination of different factors, experimental and software-related ones such as the thickness and orientation of the particle and the detection of low intensity of ordering-related diffraction spots (as discussed in details in Fig. S6, S8 and S9, respectively). Furthermore, there is also a tendency for unreliability in the phase mapping of the particle's extremities: as they are thin, the global intensity of the diffraction spots, and thus of the ordering-associated ones, fades. Despite the application of the phase reliability parameter, the remaining detection of the $Fd\bar{3}m$ phase on the phase map of the ordered *o2*-LNMO results from this last-mentioned phenomenon. It is further revealed that the zone belongs to the $P4_332$ phase rather than $Fd\bar{3}m$, as discussed in Fig. S11.

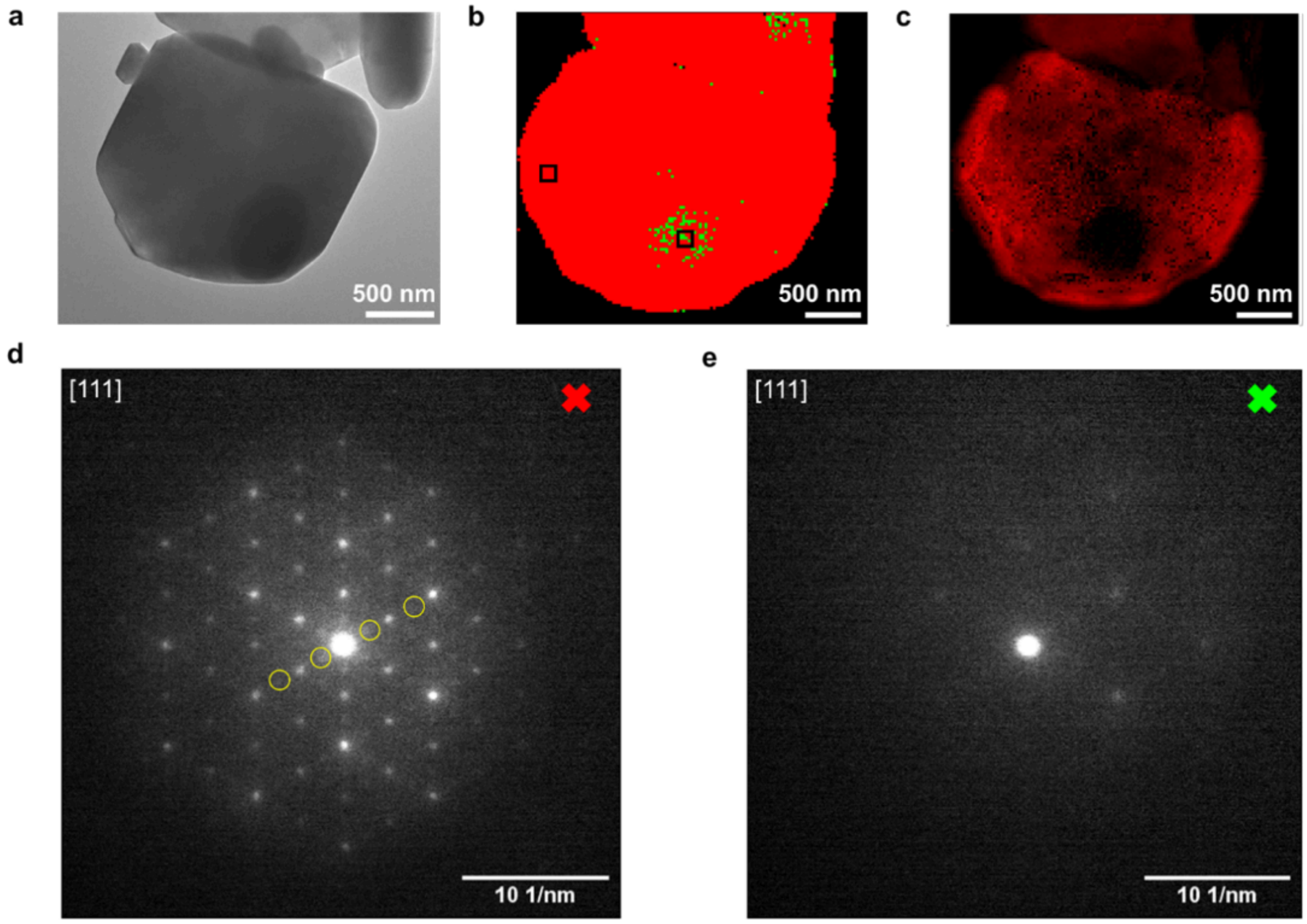

**Figure S6: Thickness induced errors in phase mapping. a** TEM image of a particle from sample *o2*-LNMO. **b** Phase map of the given particle. Red zones are detected as corresponding to the disordered $Fd\bar{3}m$ phase and the green zones to the ordered $P4_332$ phase. **c** Phase map combined with a phase reliability value ranging from 10 to 30. The resulting map indicates disordering throughout the particle, from the core to the surface. All scale bars are 500 nm in **a-c**. There is a small tilt between the TEM image and the ACOM acquisition causing orientation difference in **a** and obtained maps in **b-c**. **d** A diffraction pattern taken from the zone detected as corresponding to the $Fd\bar{3}m$ phase in **b** marked with a square in the red zone. The presence of ordering spots is visible, and yellow circles point out couple of them along $[0\bar{1}\bar{1}]$ direction as an example, showing thus that this zone detected by the software to be disordered is in fact ordered. **e** A diffraction pattern taken from the zone detected as $P4_332$ phase in **b** marked with a square in the green zone showing nearly no diffraction spots, showing thus that in that case the ordering cannot be observed.

The phase map of the particle indicated that overall the particle is described in $Fd\bar{3}m$ phase (Fig. S6c). However, the ordering-related diffraction spots observed on the diffraction pattern given in Fig. S6d indicate that the particle is in fact described in the $P4_332$ phase. It is due to diffusion effects caused by the thickness of the particle, the software cannot effectively differentiate between the intensity originating from low-intensity diffraction spots and background noise. Consequently, the pattern is indexed as the $Fd\bar{3}m$ phase. Secondly, before applying the phase reliability factor, a small zone on the particle is indeed indexed as $P4_332$ phase (Fig. S6b). As seen in the associated diffraction pattern in Fig. S6e, there are no discernable diffraction spots. As deduced from the contrast difference in the TEM image given in Fig. S6a, this zone appearing darker is thicker, possibly due to the presence of another particle underneath. The noisy patterns as observed in Fig. S6e naturally generates a better match with the theoretical pattern containing more diffraction spots, in our case the one indexed in the $P4_332$ space group, leading to an unfounded phase identification. Note that this region is later eliminated with the phase reliability criteria, as seen on Fig. S6c**. In summary, choosing a thick particle can jeopardize the accuracy of the data analysis. Therefore, manual verification is needed to ensure that the particle is adapted for the study.**

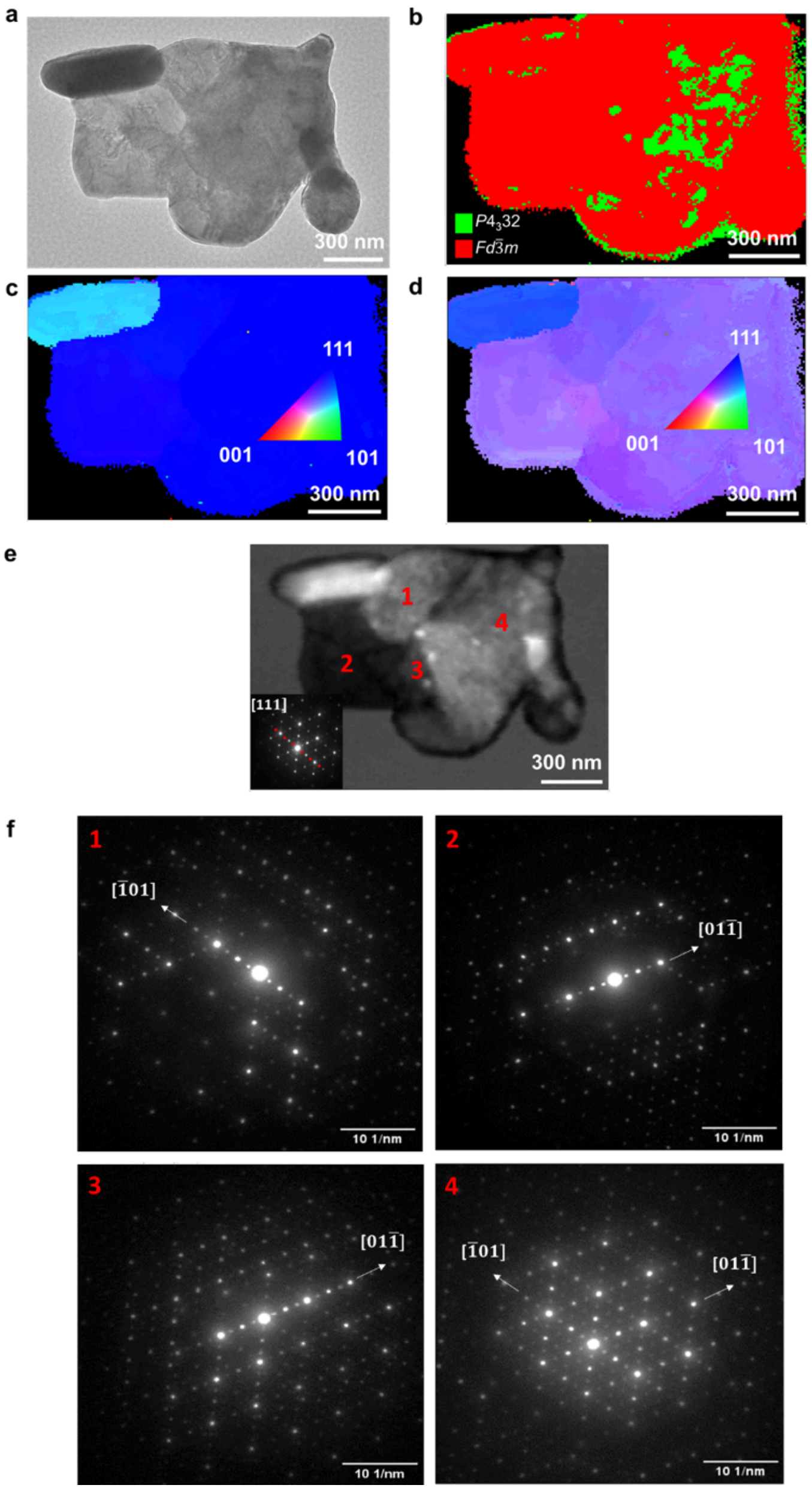

**Figure S7: Orientation induced errors in phase mapping. a** TEM image of a particle from sample *o1*-LNMO. **b** Phase map of the given particle. Red zones are detected as corresponding to the disordered $Fd\bar{3}m$ phase and green zones to the ordered $P4_332$ phase. **c** Orientation map of the particle indicating the [111] zone axis and **d** Deviation map from the zone axis. The maps show the orientation according to the color code given in inset. Nuances of the same color are observed along on the deviation map given on d indicating small orientation changes. **e** Virtual dark field image obtained from ordering diffraction spots on [$\bar{1}$01] direction of [111] zone axis (shown in inset). The absence of the selected spots leads to dark regions, while bright regions show their presence. Four regions with varying shades of gray are numbered. All scale bars are 300 nm in (a-e). **f** Electron diffraction patterns from numbered regions of (e) show tilts (up to ~4°) around the same zone axis of [111] within the same crystal.

The phase map of the chosen particle (Fig. S7a-b) indicates a predominance of disordering in the sample. However, when only the ordering spots along one axis are chosen (as demonstrated on the pattern given in inset of Fig. S7e), there are illuminated zones on the virtual dark field pattern, indicating their presence. The orientation map (Fig. S7c) obtained along the z-axis show similar orientation, whereas slight variations (~ up to 4°) of the same orientation can be observed on the deviation map (Fig. S7d). The domains corresponding to differences in orientation match exactly those observed in the virtual dark field image given in Fig. S7e. Fig. S7f further clarifies the situation; the electron diffraction patterns collected for these different regions correspond to a common overall orientation along the [111] direction, however, slightly tilted along the <110> axis. Even if still visible along the [110] direction, this tilt results in a decrease in the number of visible ordering spots available for the software to identify the domain as corresponding to an ordered $P4_332$ phase. It thus induces that the majority of the zones on the particle are incorrectly detected as the disordered $Fd\bar{3}m$ phase. This error emphasizes the importance of working along a well-defined zone axis for accurate phase matching.

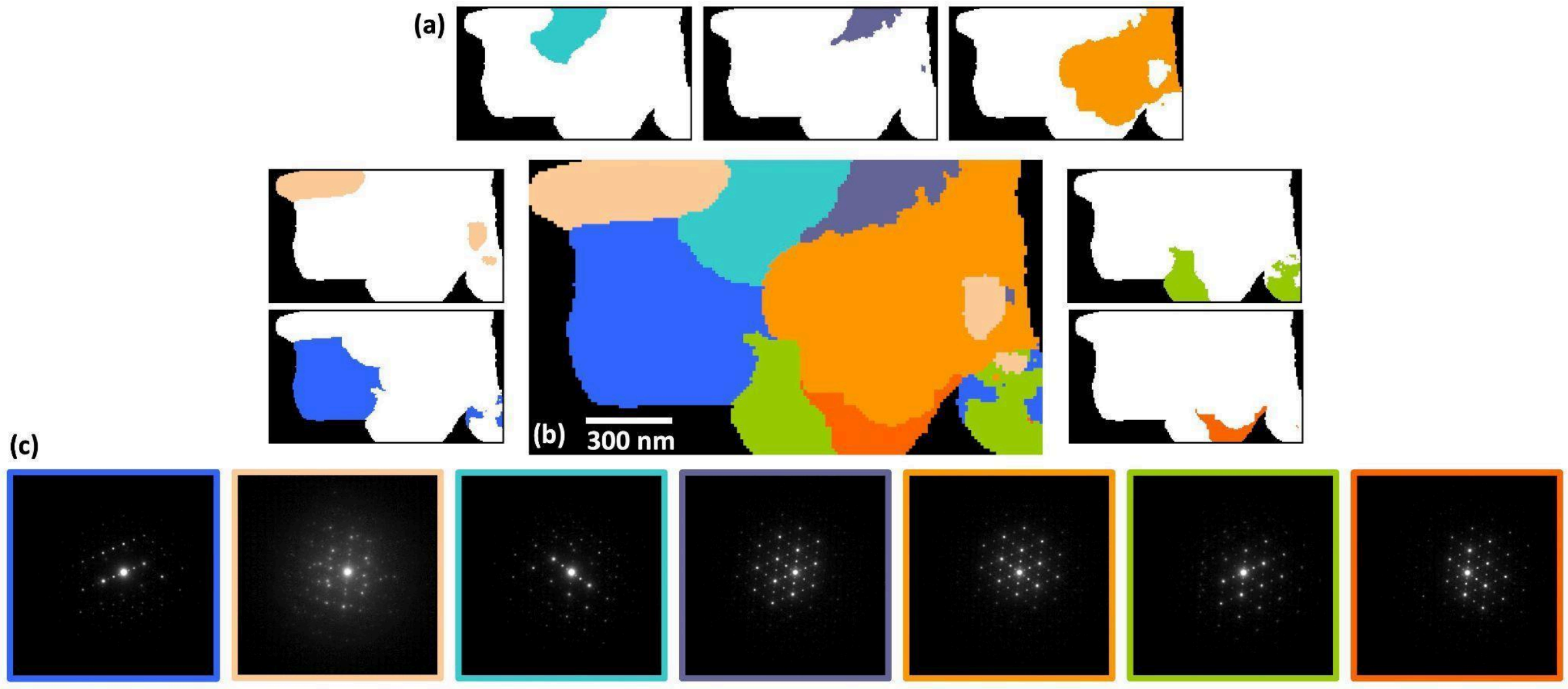

**Figure S8: Orientation mapping results and diffraction data analysis using Non-Negative Matrix Factorization (NMF) applied to 4D-STEM datasets of the LNMO particle**. (a) Segmentation maps showing distinct orientation clusters identified in the dataset. (b) Reconstructed orientation map over a 2 µm region, highlighting spatial variations in grain orientations. (c) Diffraction patterns from raw data corresponding to orientation clusters.

Fig. S8 presents the results of orientation mapping and diffraction data analysis using Non-Negative Matrix Factorization (NMF) [16] applied to 4D-STEM datasets from the *o1*-LNMO particle. This unsupervised clustering approach allowed us to identify high distinct orientation clusters within the scanned region (Fig. S8a), in addition to a background-related cluster. The spatial distribution of these clusters is reconstructed in the orientation map (Fig. S8b), revealing significant heterogeneities in local crystal orientations. These variations help explain the challenges encountered in accurately detecting ordering within the particle, as deviations from the ideal zone axis can suppress or distort the characteristic diffraction features associated with ordered structures.

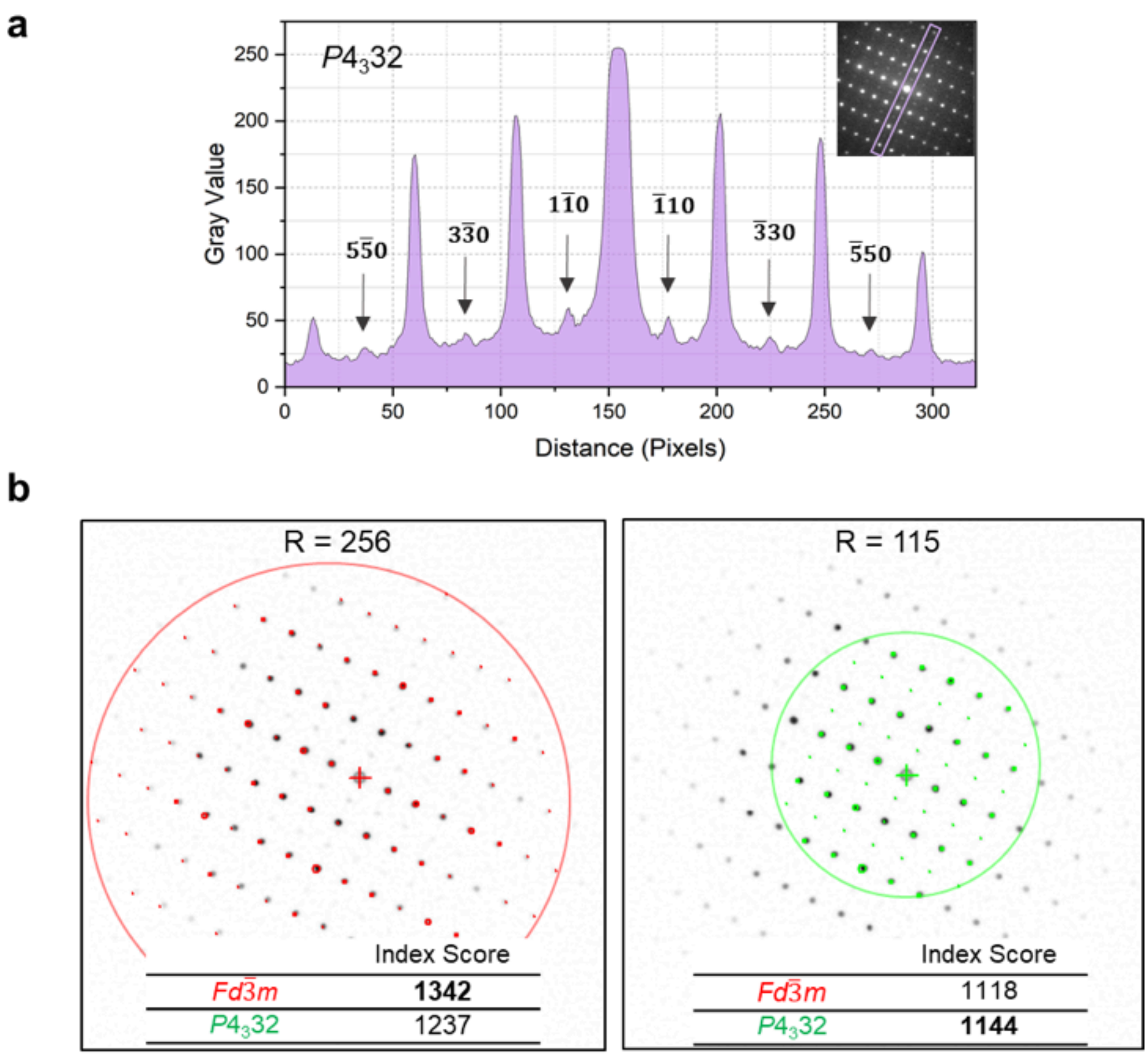

**Figure S9: Detection radius induced errors in phase mapping. a** Line integration profile on a diffraction pattern indexed in the $P4_332$ space group. The corresponding diffraction pattern along [112] direction is given in the inset. **b** Influence of the detection radius change during pattern matching. The experimental pattern is in black, and the theoretical pattern of the $Fd\bar{3}m$ phase in red and the theoretical pattern of the $P4_332$ phase in green. A larger radius selection (R=256 px on the left) leads to a better match with the $Fd\bar{3}m$ phase, while a smaller radius (R=115 px on the right) matches better with the $P4_332$ phase. The associated index scores show the shift in the assignment of the match.

Fig. S9 shows the impact of changing the detection radius on the attribution of phases. The chosen diffraction pattern is indexed as the ordered $P4_332$ phase (Fig. S9a). However, when a detection radius of 256 is selected, the pattern incorrectly matches with the $Fd\bar{3}m$ phase, as indicated by the higher indexation score (Fig. S9b). Fortunately, using a smaller radius corrects the matching error by searching only the intensity of the extra spots closer to the central beam, and the pattern is correctly indexed with the $P4_332$ phase. **This shows that when working with new materials, it is important to determine the critical radius. This requires manual testing to determine the distance from which the indexation score shifts.**

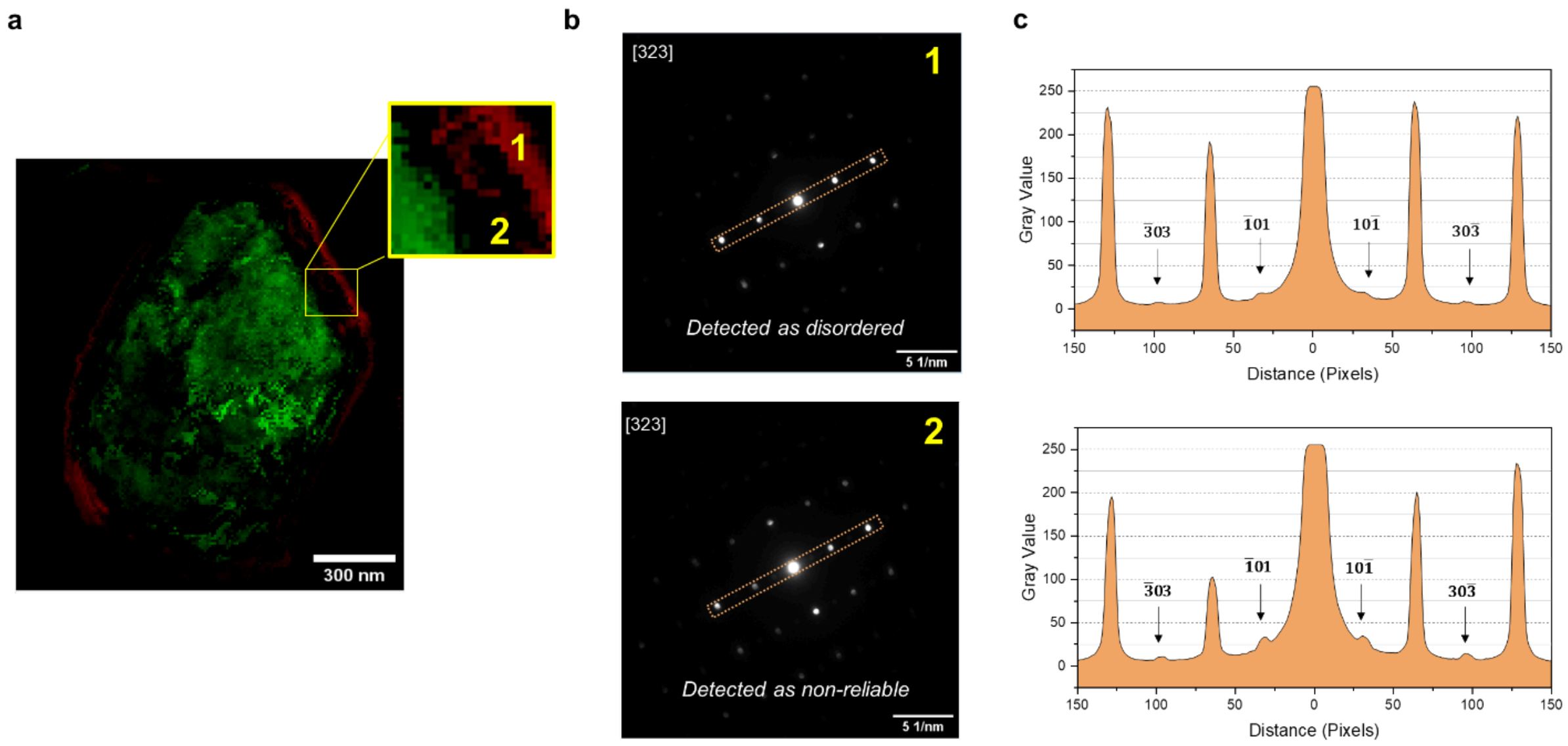

**Figure S10: Zones matched with *Fd*$\bar{3}$*m* phase in *ordered o2*-LNMO. a** Phase map combined with a phase reliability value ranging from 10 to 30 showing disordered (*Fd*$\bar{3}$*m*) phase and non-reliable zones on extremities as marked as 1 and 2 in the inset, respectively. **b** Electron diffraction patterns associated with a region detected as disordered (top) and non-reliable (bottom). **c** Line integration profiles corresponding to the 101 axis within dashed lines. The low-intensity ordering peaks are present even though they were not detected.

The phase map of the *ordered o2*-LNMO unexpectedly presents multiple non-reliable and disordered regions. When analyzing the disordered zones (Fig. S10b-c, top), there are ordered spots of low intensity (17 and 7 in gray value for $\bar{1}01$ and $\bar{3}03$ spots, respectively) along the [110] axis which falls under the inherent intensity detection limit. As a result, the pattern is falsely matched with the *Fd*$\bar{3}$*m* phase. On the other hand, when the non-reliable zones are analyzed (Fig. S10b-c, bottom), there are also ordering spots present, but the intensity is too low to be detected as the $P4_332$ phase but too high to be classified as the *Fd*$\bar{3}$*m* phase (33 and 10 in gray value for $\bar{1}01$ and $\bar{3}03$ spots, respectively). Consequently, these zones are classified as non-reliable because the indexation scores of both matches are too close. Here, the low intensity of the ordering spots is dependent on multiple factors such as the thickness of the particle, and information loss at the particles' surface due to thin edges. After careful analysis, it is validated that all zones on the studied particle of *ordered o2*-LNMO belongs to the $P4_332$ phase.

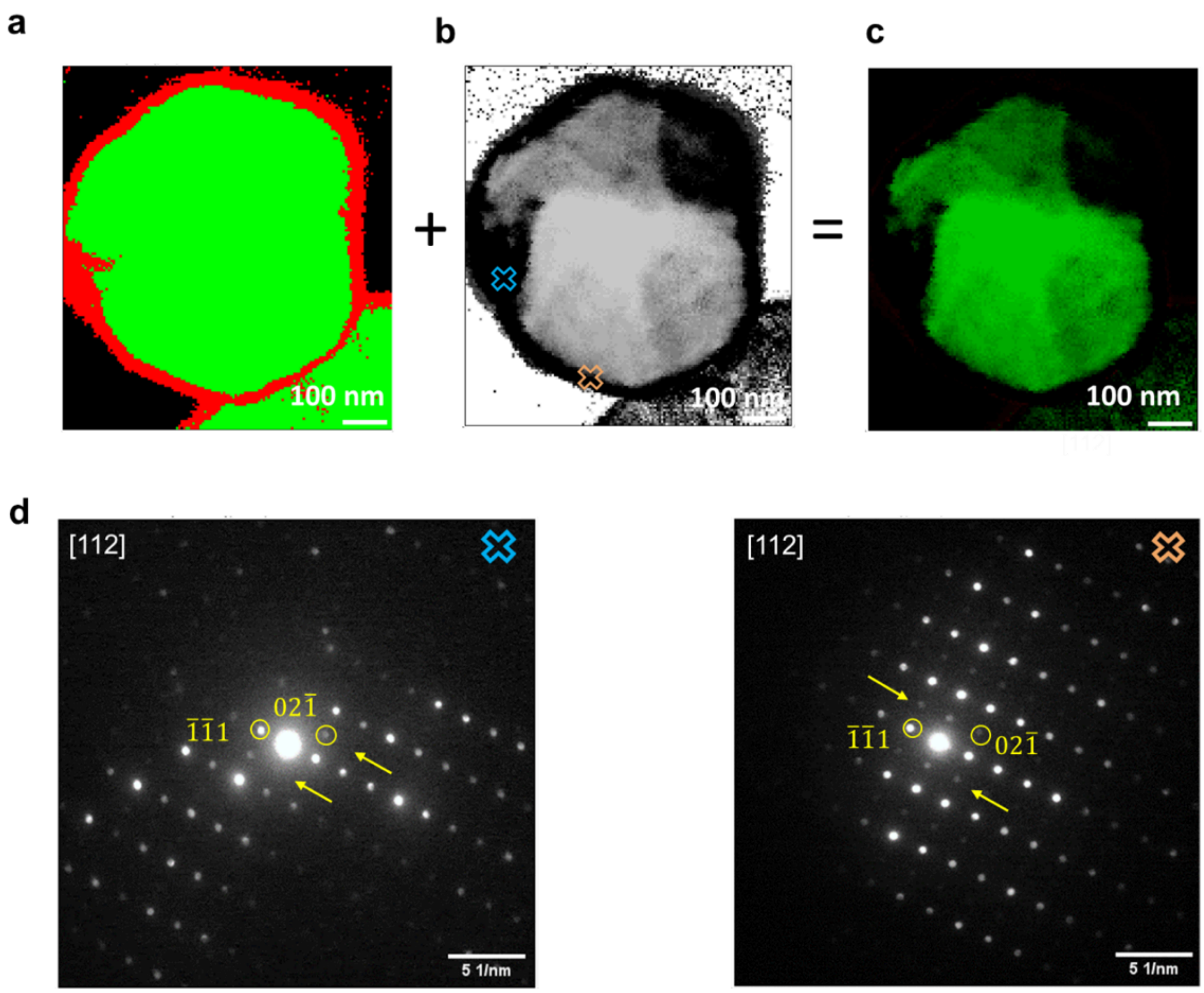

**Figure S11: Loss of information on *ordered o1*-LNMO. a** Phase map of the *ordered o1*-LNMO. Red represents the $Fd\bar{3}m$ phase, and green the $P4_332$ phase. **b** Phase reliability map within a value ranging from 10 to 30. Blue and orange crosses point out examples of regions that fell under the imposed reliability limit. **c** Combined phase map with the defined phase reliability value. All zones detected as the $Fd\bar{3}m$ phase got eliminated. Scale bars at **a-c** are in 100 nm. **d** Electron diffraction patterns of the zones identified in blue and orange in **c**. Yellow arrows point out low-intensity ordering spots.

There is a large number of areas where the ordering information is lost due to non-reliability. Here, this loss occurs due to a combination of two factors: the domain orientation as seen in Fig S8, and information loss at the extremities of the particle. In some cases, the zones are eliminated because of the tilt and the limited number of detected visible ordering spots (Fig. S11d, left). In other cases, the elimination is due to the intensity detection threshold (Fig. S11d, right). However, all collected diffraction patterns were verified and confirmed to belong to the $P4_332$ phase.

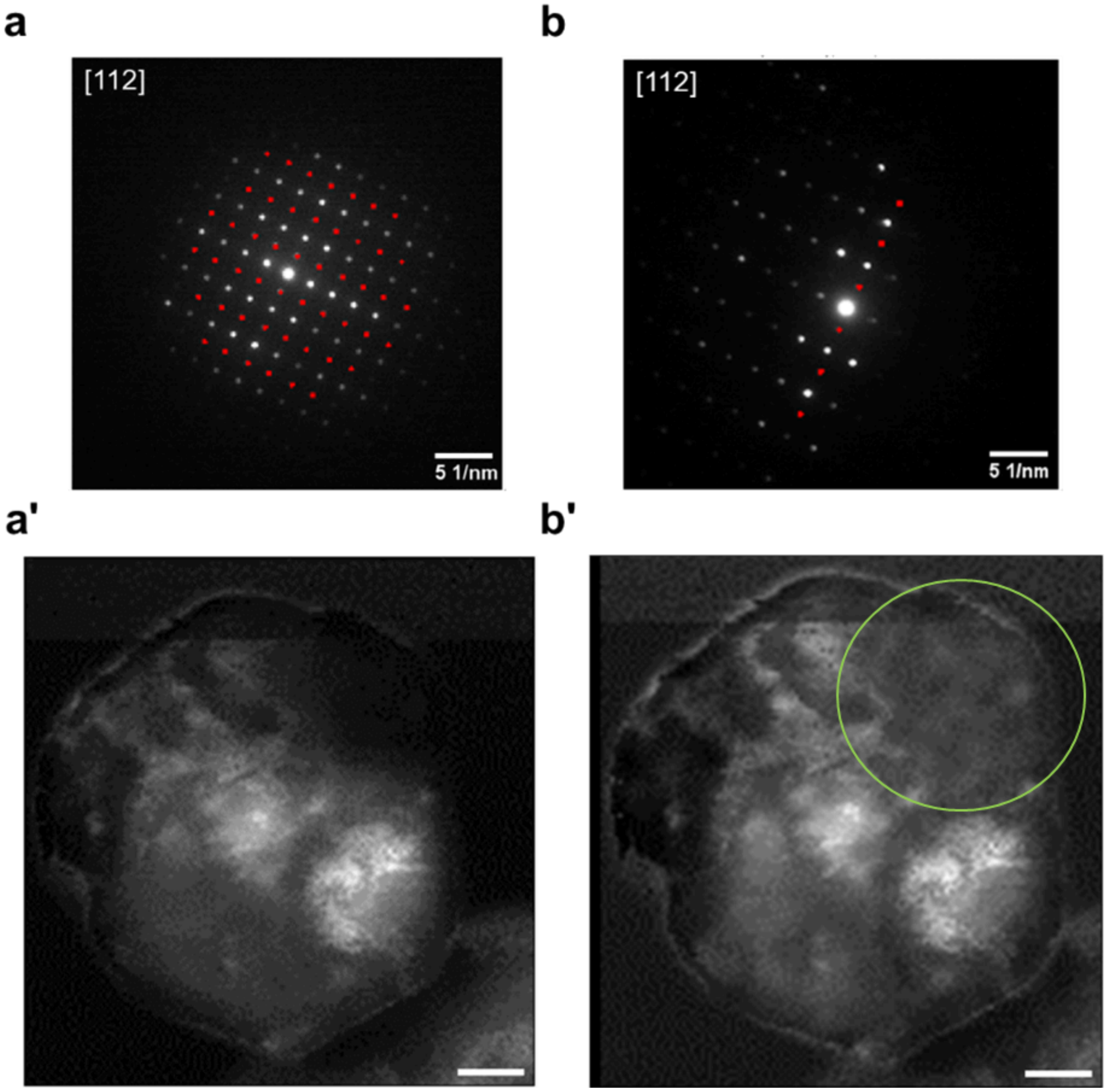

**Figure S12: Selection of the diffraction spots for intensity search. a** Selection of all the ordering diffraction spots present within the selected radius at the [112] zone axis and **a’** the associated virtual dark field map. Bright zones indicate the presence of the ordering diffraction spots, while dark zones indicate their absence. This intensity search corresponds to an « Astar-like » choice where diffraction spots are searched within a defined circle. **b** Selection of the ordering spots present on the common axis visible throughout the particle at the [112] zone axis and **b’** its associated virtual dark field map. Scale bars at **b-b**’ are in 100 nm.

When all the visible ordering diffraction spots are selected on the zone axis for intensity search, this may induce errors by orientation, as discussed in Fig. S7 and S8. By comparing the two virtual dark field maps in Fig. S11a’ and Fig. S11b’, we can observe that the intensity of the ordering spots coming from the zone marked with a green circle is only detected when only a few ordering spots are chosen. This highlights the importance of determining the common axis present throughout the particle to avoid errors caused by differences in domain orientations.

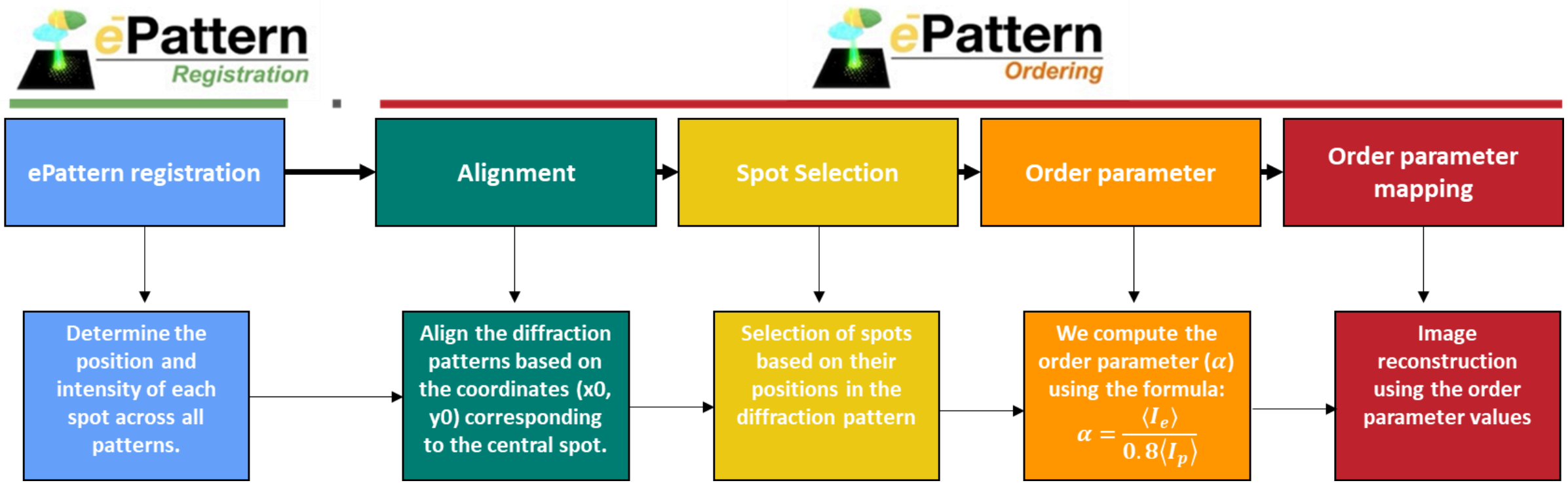

**Figure S13:** Workflow of ePattern_Ordering for the Order parameter quantification process based on ePattern_Registration process.

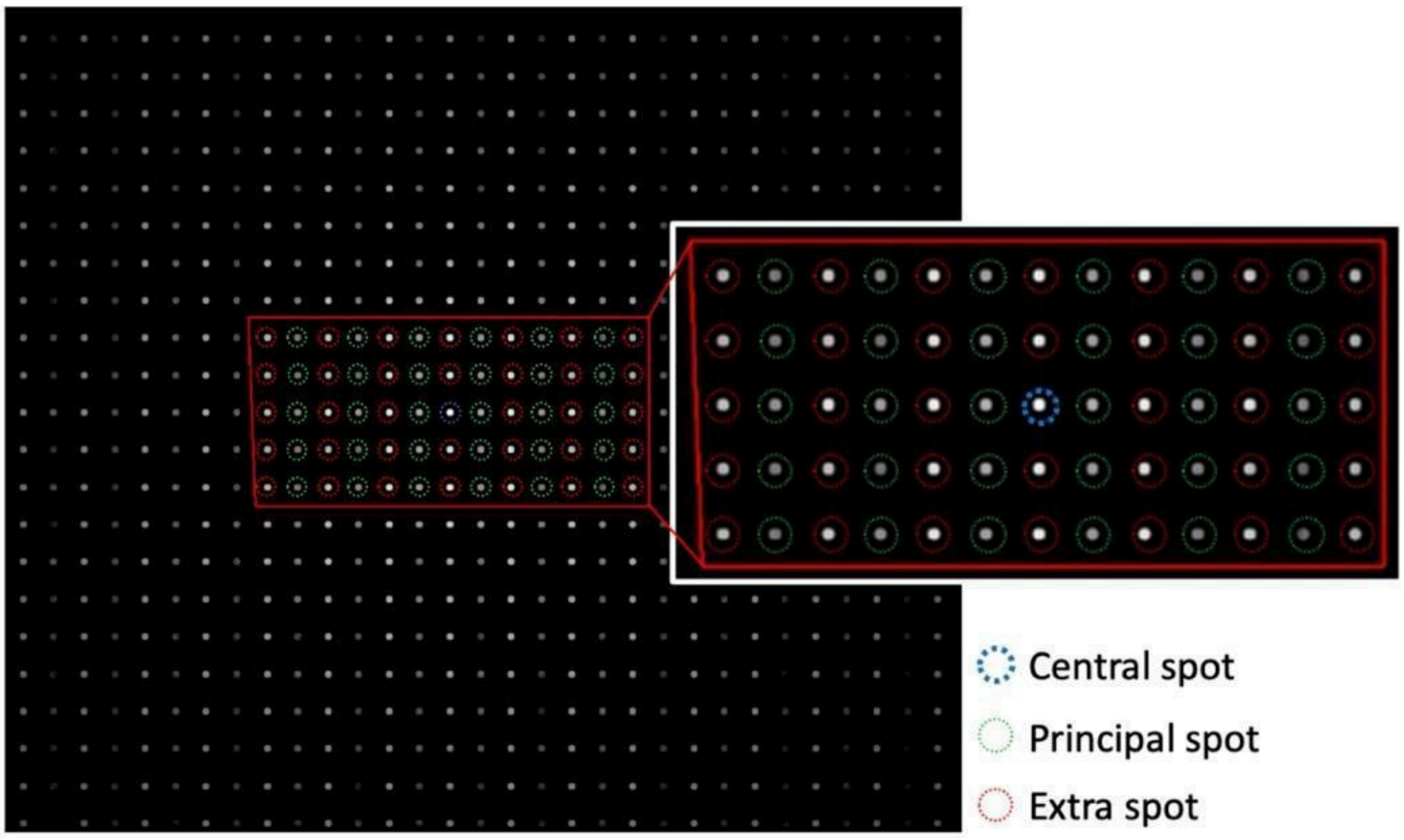

**Figure S14:** Electron diffraction pattern simulated for ordered LNMO (Order LNMO CIF file) by MacTempas2 software in order to evaluate the maximum order parameter based on intensity ratio between main and extra diffraction spots.

To establish a reference for quantifying the degree of ordering in spinel LNMO, we simulated electron diffraction patterns corresponding to a perfectly ordered structure using the MacTempasX software (version 2). MacTempasX is a widely adopted simulation platform in transmission electron microscopy, particularly suited for modeling electron diffraction patterns, high-resolution TEM images, and related electron scattering phenomena. It implements the multislice algorithm, in which the interaction between an incident electron wave and a crystalline specimen is calculated by dividing the crystal into a series of thin slices along the beam direction. At each slice, the electron wave is modulated by the projected atomic potential and then propagated to the next slice using Fourier-based Fresnel propagation. This iterative approach ultimately generates a complex exit wave at the bottom of the sample, which is Fourier transformed to yield the simulated diffraction pattern.

For this study, the simulation was performed using the crystallographic information file (CIF) corresponding to the ordered LNMO phase (space group $P4_332$), with the zone axis aligned along [110], as used in our experimental conditions. Parameters such as accelerating voltage, beam convergence, sample thickness, and Debye-Waller factors were carefully chosen to replicate the experimental setup. The resulting simulated diffraction pattern exhibits both the main diffraction spots common to disordered and ordered spinel structures, and the extra superlattice reflections characteristic of cation ordering.

These simulations provide a critical baseline for evaluating the maximum theoretical order parameter, defined as the intensity ratio between selected ordering reflections and the principal reflections. By comparing these simulated intensities with those extracted from experimental 4D-STEM data, we were able to normalize and quantify the extent of ordering at the particle level. This approach enables a direct interpretation of structural modulations due to cation ordering and facilitates discrimination between perfectly ordered, partially ordered, and disordered regions across the LNMO samples.

## Correlation coefficients

The Pearson correlation coefficient [10,11] is a statistical measure that evaluates the linear relationship between two continuous variables. It is calculated using the formula:

$$r = \frac{cov(X,Y)}{\sigma_X \sigma_Y} = \frac{\sum_{i=1}^{n} (X_i - \langle X \rangle)(Y_i - \langle Y \rangle)}{\sqrt{(X_i - \langle X \rangle)^2 (Y_i - \langle Y \rangle)^2}}$$

Where $cov(X, Y)$ is the covariance of variables X and Y, $\sigma_X$ is the standard deviation of the variable X and $\sigma_Y$ is the standard of the variable Y. The resulting value, $r$, ranges from -1 to 1, where $r$ =1 indicates a perfect positive linear relationship, $r$ = -1 indicates a perfect negative linear relationship, and $r$ =0 indicates no linear relationship. We use Pearson's correlation to identify any linear dependencies between thickness and order.

## Thickness-Ordering Relation

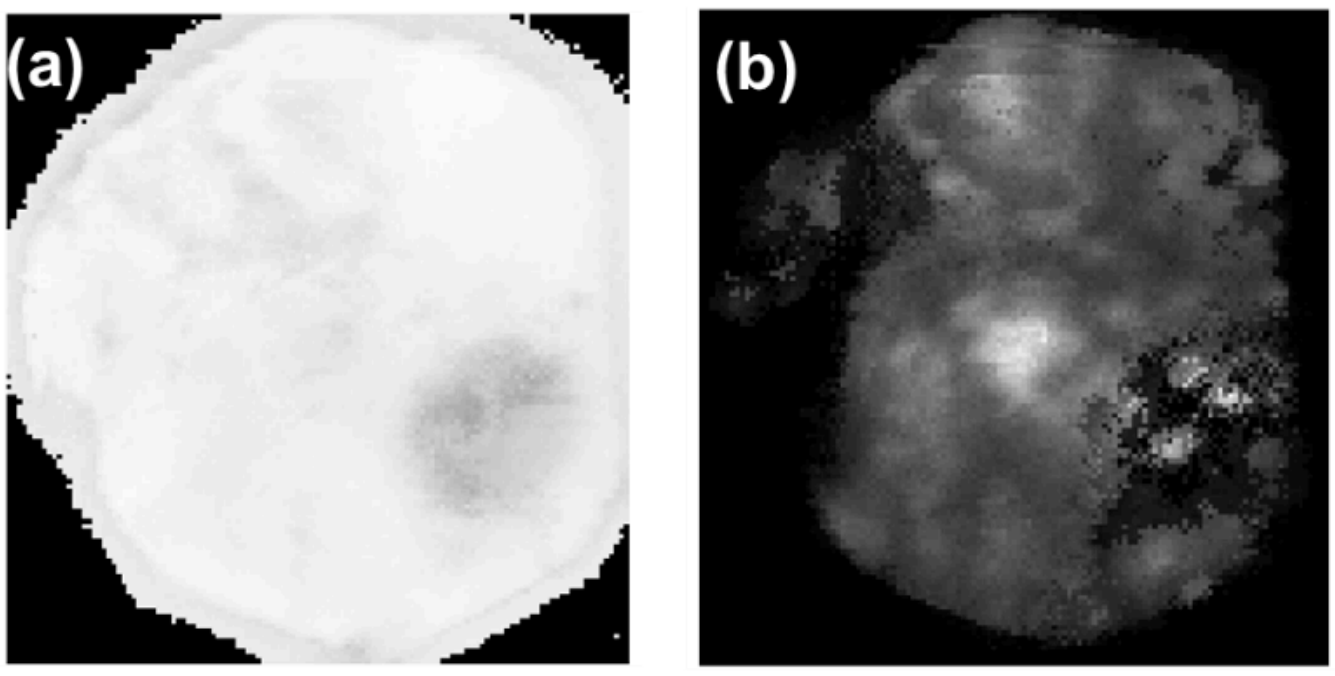

(c)

| o1-LNMO | Direct beam intensity |
| --- | --- |
| Order intensity | 0.436 |

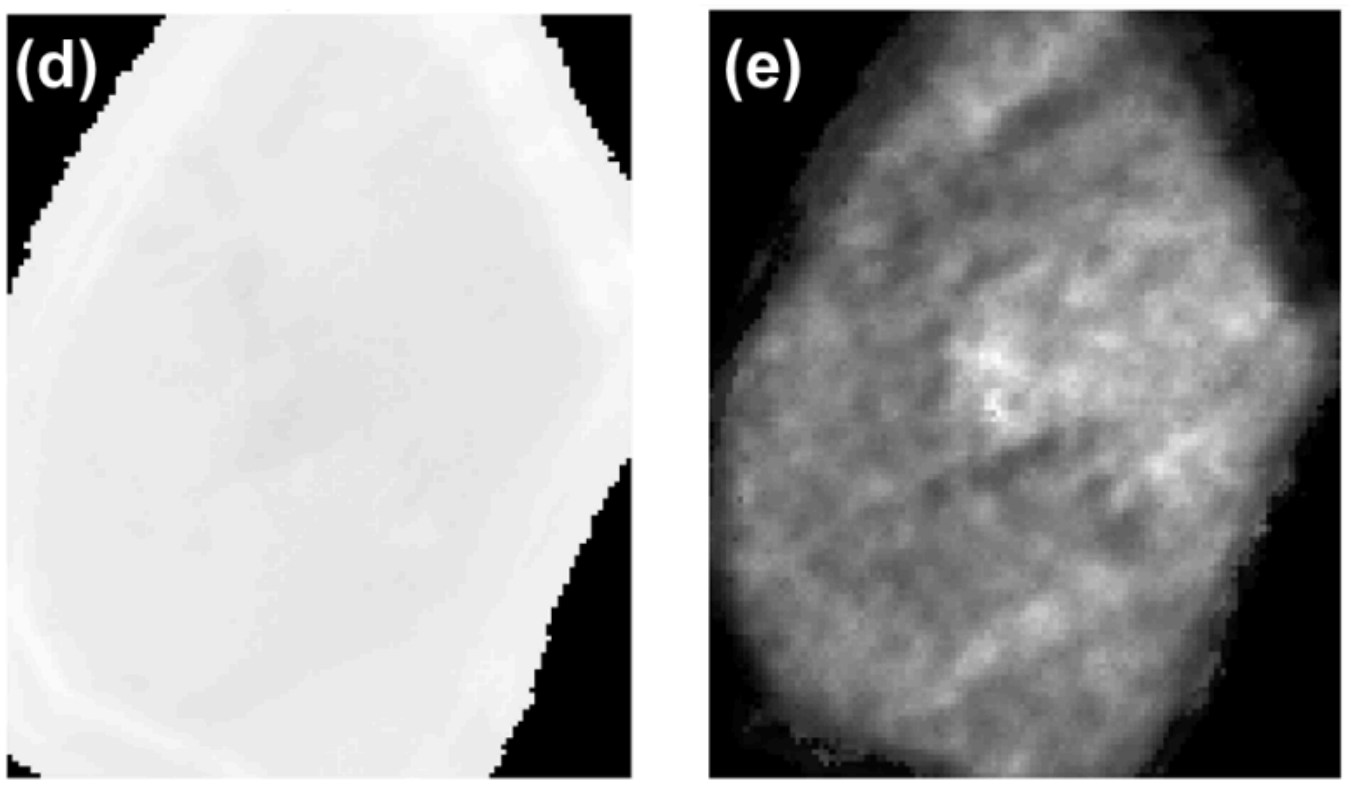

(f)

| o2-LNMO | Direct beam intensity |
| --- | --- |
| Order intensity | 0.508 |

**Figure S15. Order parameter and crystal thickness. a,d** the thickness map of the two particles *o1*-LNMO and *o2*-LNMO. **b,e** order maps for *o1*-LNMO and *o2*-LNMO.**c,f** the pearson correlation coefficient between the order and the direct beam intensity, the mean principle spots intensity and the mean ordering (extra) spot intensity.

To investigate the relationship between particle thickness and the calculated order parameters, we determined the thickness of each pixel in the sample based on the intensity of the central diffraction spot. To quantify the correlation between thickness and order, we computed the Pearson correlation coefficient, which ranges from -1 to 1. A value close to 1 indicates a strong positive correlation, while values near -1 and 0 signify strong negative and no correlation, respectively. For *o1*-LNMO, the Pearson correlation coefficient was found to be 0.436, indicating a weak correlation between order and thickness. In contrast, for *o2*-LNMO, the coefficient was 0.54, suggesting a stronger correlation. These results were expected, as the intensities were obtained under identical beam conditions. Given that the intensity of the extra diffraction spot is directly related to the intensity of the central spot, a higher correlation was anticipated.

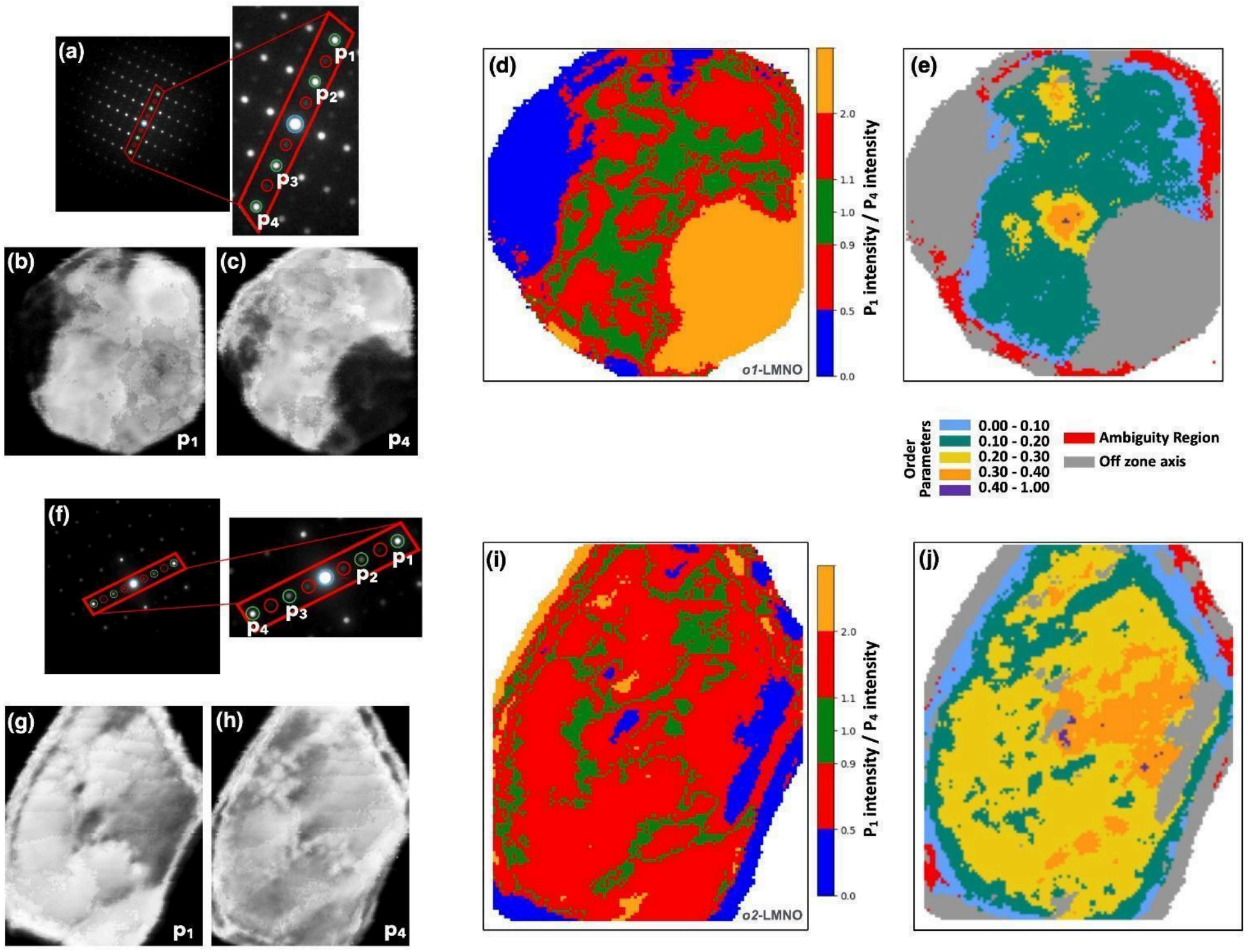

**Figure S16:** Zone axis deviation analysis based on intensity comparison of opposite diffraction spots. **a,f** diffraction patterns from the particles *o1*-LNMO and *o2*-LNMO where ordering spots (extra spots) are surrounded in red and the principle ones are surrounded in green. **b,c,g,h** virtual dark-field images obtained using the p1 spot $(\bar{4}40)$ and p4 spot p4 $(4\bar{4}0)$ for *o1*-LNMO (b,c) and p1 $(\bar{4}04)$, p4 $(40\bar{4})$ for *o2*-LNMO (g,h). **d,i** Off-zone axis map generated by dividing the intensities of (b, c) and (g,h) pixel by pixel. **e,j** Ordering map overlaid with the off-zone axis region in grey.

In electron diffraction patterns, changing intensity between opposite or symmetrical diffraction spots indicates a deviation from the zone axis. To investigate this, we generated virtual dark-field images for two opposite spots in each particle p1 and p4, Fig. S16b and S16c for *o1*-LNMO and Fig. S16g and S16h for *o2*-LNMO. To identify regions suitable for accurate order quantification, we computed pixel-by-pixel intensity ratios between the reference images, yielding the maps presented in Fig. 3d and 3i. Values approaching unity (in green) indicate nearly equal intensities, corresponding to positions close to the zone axis. We define the zone axis condition as fulfilled when the intensity ratio p1/p4 lies between 0.5 and 1. Regions deviating from this condition were identified as off-zone axis areas and overlaid in grey onto the ordering maps shown in Fig. S16e and S16j. These grey regions delineate zones where ordering spot detection is compromised due to deviation from the ideal orientation. In contrast, the non-grey areas (expect the ambiguity regions isolated due to higher level of noise hindering spot detection), remaining within the defined zone axis threshold, provide reliable conditions for the quantitative evaluation of order parameters.

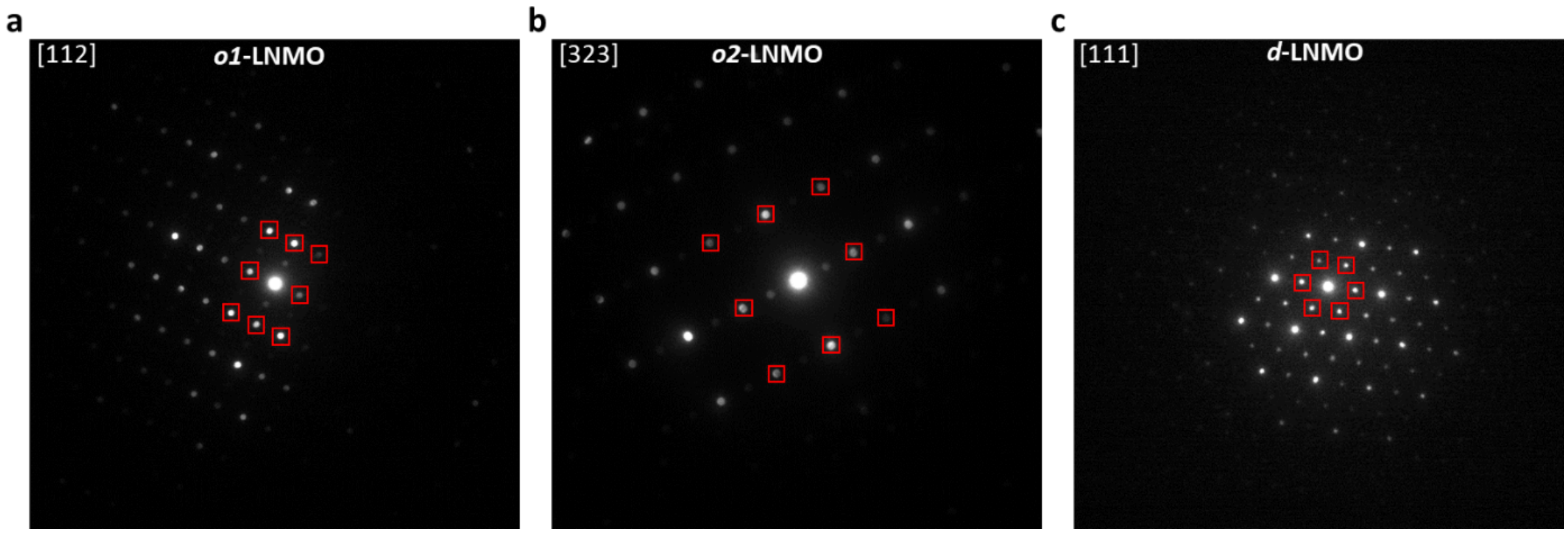

**Figure S17: (a,b,c)** Diffraction patterns acquired from samples *o1*-LNMO, *o2*-LNMO and *d*-LNMO, respectively. Red squares indicate principal spots selected for lattice parameter calculation.

For lattice parameter determination, we selected diffraction spots located closest to the central beam to minimize orientation effects and allow comparison between the three LNMO samples. For *o1*-LNMO (Fig. S17a), reflections from the {220}, {311} and {331} families were selected, for a total of eight spots. For *o2*-LNMO (Fig S17b), reflections from the {202}, {311}, and {331} families were used, also for eight diffraction spots. For *d*-LNMO (Fig S17c), six equivalent spots from the {220} family were chosen.

## Calibration process for lattice parameter calculation

To ensure precise measurements, we performed a pixel size calibration using the Py4DSTEM package, with polycrystalline gold (Au) particles as a reference material. We conducted a 4D-STEM experiment under the same experimental conditions as those used for the LNMO particles studied in this work. Before calibration, an accurate spot detection step is essential. This was achieved using the template matching process with a probe template, which was cross-correlated with the entire dataset to precisely identify the diffraction spots. Then we performed a stepwise calibration process to correct for potential distortions. Concerning our samples, we performed first the origin calibration, this step allows to correct for the shift in diffraction patterns that occurs as the sample is raster-scanned. Then we applied the ellipticity calibration which corrects distortions caused by lens aberrations, ensuring that the diffraction spots maintain their true shape. Finally, the pixel-size calibration using the known lattice parameters of gold, allowing us to convert pixel measurements into absolute distances. For a more detailed explanation of the calibration process using a known standard and spot detection, we refer to the Py4DSTEM article[15]. The specific calibration procedures we followed can be found in the Py4DSTEM notebooks available on GitHub.